\documentclass[twocolumn]{aastex631}

\usepackage{ulem}       

\newcommand{\z}{$Z$}

\newcommand{\msun}{M$_\odot$}

\usepackage{xcolor}

\definecolor{my_color}{HTML}{3a18b1}

\newcommand{\prot}{\ensuremath{P_{\mbox{\scriptsize rot}}}}
\definecolor{kcolor}{rgb}{0.54, 0.17, 0.89}

\newcommand{\kms}{\ensuremath{\mbox{km s}^{-1}}}

\newcommand{\li}{\mbox{$\rm [Li/Fe]$}}

\usepackage{appendix}
\usepackage{amsmath}

\shorttitle{rapid \prot\ at high J$_{\rm R}$}
\shortauthors{Rampalli et al.}
\graphicspath{{./}{figures/}}

\begin{document}

\title{Wrinkles in Time -- I: Rapid Rotators Found in High Eccentricity Orbits}

\correspondingauthor{Rayna Rampalli}
\email{raynarampalli@gmail.com}

\author[0000-0001-7337-5936]{Rayna Rampalli}
\altaffiliation{NSF GRFP Fellow} 
\affiliation{Department of Physics and Astronomy, Dartmouth College, Hanover, NH 03755, USA}

\author[0000-0002-3041-7822]{Amy Smock}
\affiliation{Steward Observatory, University of Arizona, Tucson, AZ 85721, USA}

\author[0000-0003-4150-841X]{Elisabeth R. Newton}
\affiliation{Department of Physics and Astronomy, Dartmouth College, Hanover, NH 03755, USA}

\author[0000-0003-2594-8052]{Kathryne J. Daniel}
\affiliation{Steward Observatory, University of Arizona, Tucson, AZ 85721, USA}
\affiliation{Center for Computational Astronomy, Flatiron Institute, New York, NY 10010, USA}

\author[0000-0002-2792-134X]{Jason L.~Curtis}
\affiliation{Department of Astronomy, Columbia University, 550 West 120th St, New York, NY 10027, USA}




\begin{abstract}

Recent space-based missions have ushered in a new era of observational astronomy, where high-cadence photometric light curves for thousands to millions of stars in the solar neighborhood can be used to test and apply stellar age-dating methods, including gyrochronology. Combined with precise kinematics, these data allow for powerful new insights into our understanding of the Milky Way's dynamical history. Using TESS data, we build a series of rotation period measurement and confirmation pipelines and test them on 1,560 stars across five benchmark samples: the Pleiades, Pisces--Eridanus, Praesepe, the Hyades, and field stars from the MEarth Project. Our pipelines' recovery rates across these groups are on average 89\%. We then apply these pipelines to 4,085 likely single stars with TESS light curves in two interesting regions of Galactic action space. We identify 141 unique, rapidly rotating stars in highly eccentric orbits in the disk, some of which appear as rotationally young as the 120-Myr-old Pleiades. Pending spectroscopic analysis to confirm their youth, this indicates these stars were subject to fast-acting dynamical phenomena, the origin of which will be investigated in later papers in this series.

\end{abstract}

\keywords{stars: ages – techniques: gyrochronology - Galaxy: kinematics and dynamics – solar neighborhood}
\accepted{October 1st, 2023}


\section{Introduction}\label{sec:intro}
With new near-all-sky astrometric, spectroscopic, and photometric observational surveys, 
we have the opportunity to 
understand the Milky Way dynamics impacting the solar neighborhood, using stellar ages as an additional probe. Gyrochronology is a robust age-dating method for younger, low-mass stars ($\lesssim 2$ Gyr, 1.3 \msun), particularly G-type stars. Upon measuring stellar rotational velocities and activity levels for stars in the Pleiades and Hyades open clusters and the Sun, \cite{Skumanich} famously found that stars slow their rotation as they age. This relationship between a star's rotation period (\prot) and its age ($t$) follows a trend described by a simple power law:
\begin{equation}
\text{\prot} \propto \text{$t$}^{0.5}.
\end{equation}

Today, one of the best ways to measure stellar rotation is using photometry to observe modulation in stellar brightness as a result of starspots rotating in and out of our view. NASA's Transiting Exoplanet Survey Satellite \citep[TESS;][]{Ricker15} provides the ability to make such measurements for hundreds of thousands of nearby, bright stars. These data pair well with astrometry from the Gaia mission \citep{gaiamission} and spectroscopy from surveys like GALAH \citep{galah_mission} and APOGEE \citep{apogee} of stars in the local volume. 

This paper is the first in a series where we focus on young stars in highly eccentric Galactic orbits. It is generally assumed that stars are born on near-circular orbits. Over the lifetime of a star, its orbit can be dynamically perturbed to a more eccentric state (e.g., \citealt{Sellwood14} and references therein). We can explore exceptions to this trend with gyrochronological ages and precise kinematics. 

\subsection{More on Gyrochronology and its Caveats}

Since 1972, many research groups have used observations of stellar populations with known ages to empirically calibrate the gyrochronology age--rotation relation. The power-law exponent is typically found to be 0.50--0.65, close to Skumanich's original 0.50 value \citep{Barnes2007, Mamajek2008, Meibom09, Angus15,douglas2019, stardate, curtisrup14720}. Ultimately, this relation is intended to be inverted to obtain precise and reliable ages for field stars of unknown ages.

Gyrochronology is not applicable for all lower mass stars of all ages. As low mass stars are interacting with their circumstellar disks and contracting onto the main sequence, they start at a range of \prot. They eventually converge to a single \prot\ for a given age and temperature \citep{Barnes2007}, with the timescale for convergence being mass-dependent. Gyrochronology is only applicable after this convergence occurs. At $\approx 700$ Myr, K~dwarfs temporarily stall and only resume spinning down at 2~Gyr \citep{ngc6811, curtisrup14720}. M~dwarfs have different dynamos from their G- and K-type counterparts and likely do not follow the same power law for age-dating \citep{Newton16, Rebull18,Popinchalk21}.

Thus, using gyrochronology on solely G~dwarfs is advantageous due to their favorable timescales for convergence ($\approx 100$ Myr, see \citealt{rebull2016-1, blanco1,Boyle22}), little to no stalling, and the relative abundance and brightness of these stars. This is feasibly done with TESS. 
TESS follows in the footsteps of Kepler \citep{borucki2010} and K2 \citep{howell}, providing high-cadence, high-precision photometric data which can be used to measure \prot\ for large numbers of stars. Large-scale \prot\ measurements (Ruth Angus; private communication) and age--rotation relation calibration efforts \citep[e.g.,][]{Kounkel_2022} using TESS 
have already begun and will be incorporated into future analysis. 

It is worth noting that the short $\sim$27-day baseline for most TESS observations limits the longest \prot\ we can reliably measure in this study at 13 days (determined in Appendix \ref{sec:injection}), setting the upper age-limit $\lessapprox 1.3$ Gyr for stars $\lessapprox 1.3$ \msun\ according to age--rotation relations. 

\subsection{Interpreting Galactic Action Space}\label{sec:GAS}

In this paper, we are interested in identifying young stars in orbits that are typically associated with older populations, namely high eccentricity orbits. Rather than reconstructing orbits using each star's 6D position and velocity coordinates, we approximate their orbital size and eccentricity by projecting these kinematics into Galactocentric actions. Action space is a natural space when analyzing orbital properties and can be calculated by assuming a reasonable gravitational potential for stars in the local volume \citep[e.g.,][]{BinneyTremaine,Trick19}.

Assuming a cylindrical symmetry for disk kinematics with the $z$-axis oriented perpendicular to the plane of the disk, azimuthal action, $J_\phi$, is equal to orbital angular momentum in the $z$-direction, $L_z$. Stars with higher values for $J_\phi$ will also have larger mean orbital radii and thus larger orbits. Radial action, $J_R$, is related to orbital eccentricity. In a disk with a flat rotation curve, the radial action can be approximated by $J_{R} \propto E_R/\kappa$, where $\kappa$ is the epicyclic frequency and $E_R$ is the energy associated with the orbit's non-circular motion. In this approximation the value of $E_R$ is $E_R{=}E{-}E_c$, where $E$ is the total orbital energy, $E_c$ is the energy for a circular orbit, and each is evaluated at the angular momentum $L_z$ for that orbit. For the remainder of this work we will use $L_z$ and $J_R$.

We calculated the radial and azimuthal action for each star in our catalog following \cite{Trick19} using the \texttt{Python}-based galaxy modeling package \texttt{galpy v1.7} \citep{Bovy15}.\footnote{The galpy package can be accessed using the following link: http://github.com/jobovy/galpy.} For these calculations, we approximate the underlying potential using the default settings for the \texttt{MWPotential14} and use the St{\"a}ckle-Fudge approximation
\citep{stackelfudge}. Our orbital analysis is described in section \ref{sec:actioncalc}.

\begin{figure*}
\centering
\includegraphics[width=\textwidth]{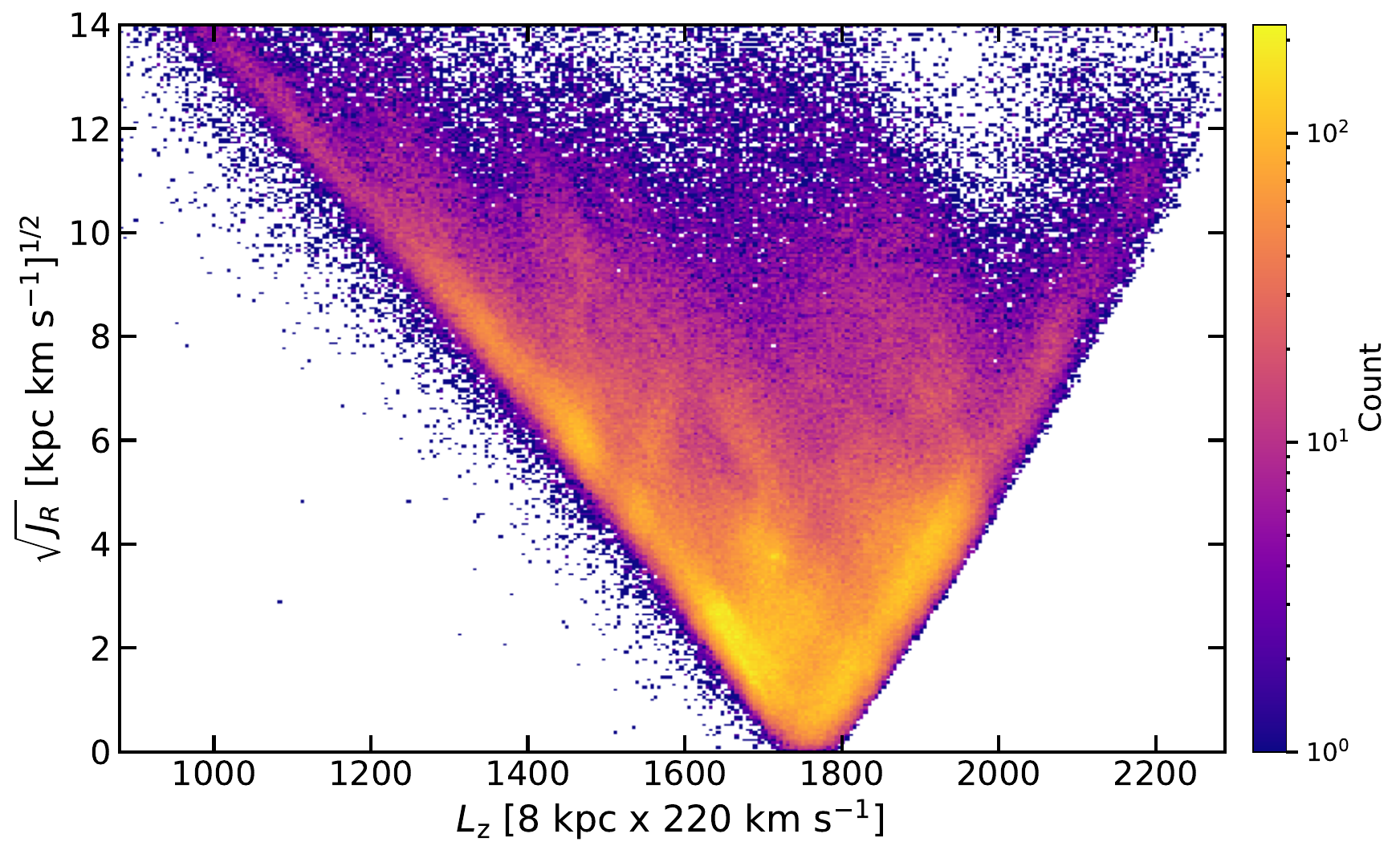}
\caption{$L_{z}$ versus $\sqrt{J_{R}}$ using updated kinematics from Gaia DR3, GALAH DR3, and APOGEE DR16. Larger $L_{z}$ is associated with a larger orbital size in the solar neighborhood, and larger $J_{R}$ is associated with a more eccentric orbit.}
\label{fig:actions}
\end{figure*}

The distribution of actions for stars in our catalog within 200~pc of the Sun are shown in Figure~\ref{fig:actions}; our catalog uses data from Gaia DR3, GALAH DR3, and APOGEE DR16 (see section \ref{sec:data}). The horizontal axis shows orbital angular momentum $L_z$, where orbits with larger mean orbital sizes will be found to the right of this plot. The vertical axis shows the square root in radial action, $\sqrt{J_R}$. 
The distribution in the Solar Neighborhood broadly fills in an inverted triangle. Stars in circular orbits (low values for $\sqrt{J_R}$) have angular momentum close to the value of the Sun's $L_{z,\odot}$, located at the lower point of the triangular distribution. Stars on highly eccentric orbits (high values for $\sqrt{J_R}$) may be interlopers in the Solar Neighborhood from other parts of the disk and thus may have angular momenta much smaller (or larger) than that of the Sun and will appear in the upper left (or upper right) of Figure~\ref{fig:actions}.

While an equilibrium distribution free of dynamical perturbations would appear smooth, \cite{Trick19} first identified the diagonal, elongated substructures in the $L_z{-}J_R$ plane at high values of $J_{R}$, which we refer to hereafter to as ``wrinkles''. These are also sometimes referred to as corrugations or simply kinematic groups, and they motivate our age investigation of stars in this space.


\subsection{This Work}

In this work, we build pipelines to measure \prot\ and test them on stars of known \prot\ to assess their performance. We then apply these pipelines to regions in high eccentricity phase-space. 
In section \ref{sec:data}, we discuss the light curves used to measure \prot, the literature we test our \prot\ pipelines on, and the datasets we use to estimate actions for stars. In section \ref{sec:methods}, we outline the design of our \prot\ pipelines. We compare our \prot\ measurements to literature measurements from five benchmark samples in section \ref{sec:results}. In section \ref{sec:actionresults} we show a preliminary \prot\ distribution for stars in a selection of high eccentricity orbits. We conclude in section \ref{sec:future} with a summary of this work and discussion about future plans for this paper series.

\section{Data} \label{sec:data}
\subsection{Measuring \prot\ with TESS Photometry}
We use TESS photometry to measure \prot, which is the foundation for determining gyrochronological ages. TESS is a nearly all-sky survey searching the nearest and brightest stars for transiting exoplanets \citep{Ricker15}. Each observational period, or sector, lasts $\approx 27$ days for a field of view that is 24 by 96 degrees. The mission obtained two-minute cadence photometry for 200,000 stars that were of high priority for its exoplanet science goals and supplied full-frame images (FFIs) from each sector at a cadence of 30 minutes. Now in its extended mission, it is continuing to provide observations based on guest investigator proposals and supply FFIs with a cadence of 200 seconds. We start with photometry from the FFIs, since the shorter cadence data is limited to certain targets. If the star does not have a light curve from the FFI-reduction pipeline we use, then we check the short cadence data. 

All of the stars observed by TESS are listed in the TESS Input Catalog (TIC) with their respective stellar parameters based on the second data release from Gaia \citep{gaiadr2} and have an associated TIC identification number \citep{TIC}. We used data accessed January 2023, which includes all sectors through sector 57.

Instrument systematics can affect the TESS photometry. This includes pointing jitter, the spacecraft's 14 day orbit, crowding from neighboring stars due to the cameras' large pixels, momentum dumps, and scattered light from the Earth and Moon (e.g. \citealt{JenkinsPA2020,stumpe_kepler_2012,Smith12}). Various teams have developed light curve reduction pipelines to mitigate these effects, including the reduction pipeline from the mission itself, TESS Science Processing Operations Center (TESS-SPOC, \citealt{tess-spoc}).

TESS-SPOC light curves are generated from the FFIs for all stars with a TESS magnitude ($T_{\rm{mag}}$) $\leq 13.5$ using simple aperture photometry (SAP). Here, the pixels are summed within an aperture optimized for signal-to-noise of the particular target star \citep{Morris17}. The resulting SAP light curve is then corrected with a set of background pixels. Given the instrument systematics discussed above, additional corrections are made by modeling and removing time-dependent instrumental signatures. Corrections to the light curve are also made to account for crowding from other stars. These light curves are known as the presearch data conditioning SAP light curves, or PDC-SAP light curves. In this analysis, we measure \prot\ using the PDC-SAP light curves from the TESS-SPOC pipeline as the default. If the star does not have a TESS-SPOC light curve, we also check for SPOC light curves, which are generated the same way but for the two-minute cadence targets \citep{spoc}; these stars can have $T_{\rm{mag}} \leq 13.5$. 


\subsection{Literature \prot\ Measurements from Kepler, K2, and MEarth} \label{sec:existing prot}
We build and test our \prot-measurement pipelines on samples of stars with previously measured \prot. While we are interested in age-dating young stars (likely 0.5--1 Gyr), high-eccentricity phase-space is predominantly made up of field stars (see section \ref{sec:actioncalc}) with \prot\ too long to measure with TESS. We choose a mix of young stars and old field stars with known \prot\ to ensure our pipelines can robustly measure the \prot\ they are supposed to and do not detect \prot\ when they should not. We crossmatch each of the \prot\ catalogs used with the TIC and Gaia DR3 to obtain 
information such as the TESS contamination ratio and Gaia astrometric information that may indicate the star is a binary (section \ref{sec:binaries}). 

For young stars, we look at four associations, three of which are open clusters of known ages. This includes 120-Myr-old Pleiades \citep{rebull2016-1}, 670-Myr-old Praesepe \citep{Rampalli21}, and 730-Myr-old Hyades \citep{douglas2019}. The community has studied rotation in these clusters for more than a decade using observations with longer baselines (particularly with the K2 mission, which had an 80 day baseline and conducted multiple campaigns covering Praesepe and Hyades), making these good laboratories in which to test \prot\ recovery. 

We also test our pipeline on stars in the Pleiades-aged Pisces--Eridanus stellar stream \citep[][called Meingast~1 by \citealt{Ratzenbock2020}]{meingast2019}. The literature \prot\ from \cite{PisEri} were derived from light curves generated from TESS FFI using simple aperture photometry and measured by visually confirming Lomb--Scargle periodograms. While our analysis also utilizes TESS data, the light curves we use are generated and confirmed with a different approach. 

For the field star population, we use the MEarth M~dwarf sample which contains a broad range of rapid and slow rotators \citep{Newton16,Newton18}. This catalog provides a sample where some periods are known to be too long for rotational modulation to be apparent over 27 days. We use stars with the most reliable signatures of rotation, designated in those works as ``grade a'' or ``grade b'' rotators. 

We list further information on the benchmark samples used in Table~\ref{tbl:clusters}.

\begin{deluxetable}{ccccc}
\centering 
\tabletypesize{\footnotesize} 
\tablecolumns{4}
\tablewidth{0pt}
\tablecaption{Literature \prot\ catalogs tested with our \prot\ pipeline}
\label{tbl:clusters}
\tablehead{
\colhead{Group} & 
\colhead{Age (Myr)} & 
\colhead {Distance (pc)} &
\colhead{\# of stars} & 
\colhead{Survey}
}
\startdata
Pleiades &	120 & 136 & 739 & K2 \\
Pisces--Eridanus & 120 & 80--226 & 101 & TESS\\
Praesepe & 670 & 186 & 1013 & K2 \\
Hyades & 730 & 48 & 247 & K2 \\
MEarth & field & $<$30 & 586 & MEarth
\enddata
\end{deluxetable}

\subsection{Building a Kinematics Catalog with Gaia, GALAH, and APOGEE}\label{s:catalog}
The Gaia mission was launched in 2013 with the goal of creating the largest and most precise 3D catalog of objects in the sky \citep{gaiamission}. It is equipped with three instruments: the astrometry instrument (ASTRO) to measure positions, the photometric instrument (BP/RP) to measure luminosities, and the radial velocity spectrometer (RVS) to determine line-of-sight radial velocities. The third and latest data release, DR3, provides full astrometric solutions for over 1.46 billion stars in the Galaxy and radial velocities for 33 million stars \citep{GaiaDr3}. 

While Gaia has spectroscopic capabilities, the resolution of RVS is not very high ($R$=11,000) given that spectroscopy is not a primary science goal of the mission. Furthermore, the precession of the Gaia spacecraft prevents the measurement of truly precise RVs with an average RV precision of 1.3~\kms\ \citep{gaia_rvs}. However, a number of ground-based surveys provide accompanying higher-resolution spectroscopic observations for stars Gaia has observed. Two such efforts are the GALactic Archaeology with HERMES, or GALAH, Survey ($R$ = 28,000) and Apache Point Observatory Galactic Evolution Experiment (APOGEE; $R$= 22,500); these both have RV precisions of around 0.1~\kms\ \citep{galah_rvs,apogee_rvs}. GALAH observes stars of all ages and locations in the Milky Way for Galactic archaeology purposes \citep{galah_mission} while APOGEE is a near-infrared survey designed to primarily measure spectra for red giant stars in the inner disk and bulge \citep{apogee}. When available, we use the more precise radial velocities from GALAH DR3 \citep{galah} and APOGEE DR16 \citep{Jonsson2020}.

We recreate the action space plot from \cite{Trick19} based on
the most recent phase-space information for 3.9 million stars observed by the Gaia mission \citep{edr3,GaiaDr3,gaia_rvs}, which we supplement with RVs from GALAH and APOGEE.
This is shown in Figure \ref{fig:actions}. We first query Gaia DR3 for all stars within 200 pc that have a measured RV and parallax errors less than 10\%, yielding 731,961 stars. We then crossmatch these results with GALAH RVs and APOGEE RVs, limiting stars with APOGEE RVs to an RV signal to noise ratio $> 10$ as suggested by the survey. This resulted in 16,937 and 47,432 crossmatches respectively. For stars that have RVs in each survey, we calculate the zero-point offset among surveys using GALAH as our standard. We find an offset of $-0.1 \pm 5.9$~\kms between Gaia and GALAH and $-0.5 \pm 6.8$~\kms\ between APOGEE and GALAH, which is reasonable given the errors associated with each survey \footnote{We also consider RVs from the Gaia-ESO, LAMOST, and RAVE surveys \citep{gaiaeso,LAMOST,RAVE}. However, the calculated offsets from Gaia, GALAH, or APOGEE exceeded the typical reported errors, so we choose not to include them in our action calculations.}. If the star does not have a GALAH RV, we use the APOGEE RV, otherwise we use the Gaia RV.

\subsection{Action Space Calculations}\label{sec:actioncalc}
 
As outlined by \citet{Trick19}, to calculate actions for our 731,961 stars, we convert the 6D observed kinematics to Galactic heliocentric Cartesian coordinates (X,Y,Z) and the corresponding heliocentric velocities (UHC, VHC, WHC) using coordinate transformations in \texttt{galpy} \citep{galpy}. U defines the radial direction toward the Galactic center, V defines the rotational direction of the Galaxy, and W defines the vertical direction toward the Galactic North pole. We incorporate the Sun's motion with respect to the Local Standard of Rest (LSR), (U, V, W) =
(11.1, 12.24, 7.25)~\kms\ \citep{Schonrich10}. Then we convert (UHC, VHC, WHC) to (ULSR, VLSR, WLSR) assuming the Sun's distance to the Galactic center is 8~kpc, its circular velocity is 220~\kms, and its height above the Galactic plane is 
25 pc \citep{galpy,Bovy12,Juric08}. We transform (X, Y, Z, ULSR, VLSR, WLSR) to Galactocentric cylindricial coordinates: (R, $\phi$, z, vR, vT, vz). Assuming the St{\"a}ckle-Fudge Galactic potential \citep{stackelfudge}, we find each star's orbital actions using \texttt{galpy}. We remove any stars with a $\sqrt{J_{R}} > 14.4$~kpc\,km\,s$^{-1/2}$ as these are likely stars in the halo. 

We simulate the propagation of parallax and RV errors in our action calculations for 1000 random stars. We find that 10\% error on parallax has minimal effect ($\leq 1\%$) on action calculations. Radial actions begin to significantly change for RV errors 5--10~\kms. The average change in $\sqrt{J_R}$ was always $ < 0.1$~kpc\,km\,s$^{-1/2}$, but the standard deviations were $\geq 0.7$~kpc\,km\,s$^{-1/2}$ for RV errors 8--10~\kms. Based on the scales of wrinkles, we choose stars with a RV error $\leq 7$~\kms, which gives an average change in $\sqrt{J_R}$ of $0.07 \pm 0.6$~kpc\,km\,s$^{-1/2}$ \footnote{Making a cut on RV error also can remove binaries, which is beneficial for gyrochronology. However, it can remove fainter rapid rotators as well (e.g., see footnote 4 in \citealt{Rampalli21}). This will not be a significant issue here since the stars are nearby, and the RV error cut is still quite high.}. Angular momentum changes are small ($< 1\%$).

\begin{figure*}
\centering
\includegraphics[width=0.65\textwidth]{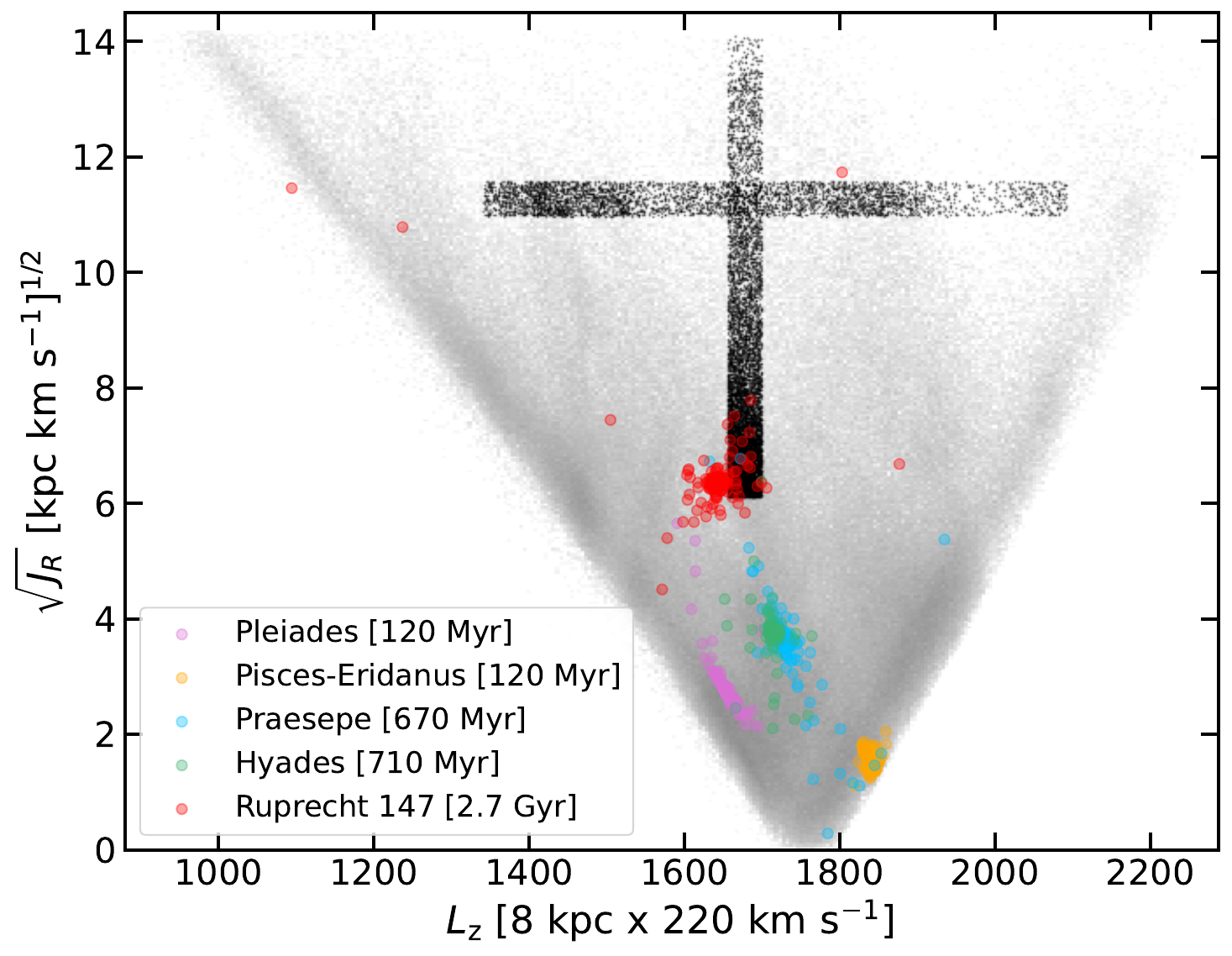}
\caption{Updated action plot as shown in Figure \ref{fig:actions} and described in \S\ref{sec:GAS} with the Pleiades (in pink), Pisces--Eridanus (orange), Praesepe (blue), Hyades (green), and Ruprecht 147 (red) overplotted. Vertical and horizontal slices chosen for \prot\ study noted in black. As the clusters increase in age, so do their radial actions, motivating our choice for studying \prot\ and finding young stars at high $J_{R}$.}
\label{fig:actionplotwstuff}
\end{figure*}

In subsequent works we will explore how combining ages and kinematics can significantly deepen our interpretation of kinematic structures in action space. For this work, we choose a vertical and a horizontal slice in high $J_{R}$ on which to run our \prot\ pipelines. The locations of these was in part motivated by where benchmark clusters fall in action space. In Figure \ref{fig:actionplotwstuff} we see the younger clusters like the Pleiades, Pisces--Eridanus, the Hyades, and Praesepe are at lower $J_{R}$, while the 2.7-Gyr-old Ruprecht~147 cluster sits a bit higher \citep{curtisrup14720}. Note that Ruprecht 147, which is not one of the clusters we study in this work, is at a distance of 302~pc, beyond our 200~pc limit. We have plotted its members (using membership cuts suggested by \citealt{curtisrup14720}) for demonstration of where an older group of stars falls in action space. We also note that there is an elongation in most of the clusters. While exploring the reason for this morphology is beyond the scope of this work, we posit this could be the result of tidal effects (e.g. \citealt{Roser19_H,Roser19_P}), the presence of binaries and their effects on the measured RVs, or interloping non-members. However, because we use rotation-based catalogs, the latter is unlikely. 

Both slices were selected using \texttt{Glue} \citep{glue}. \texttt{Glue} is an interface that allows ``linked-view" visualizations of astronomical image or catalog data. For example, users can draw an arbitrary shape on a plot with an associated table of data that will filter the particular data lying within the drawn boundaries. We recreate Figure \ref{fig:actions} in the \texttt{Glue} interface and manually highlight the two slices, which in turn creates a sub-table with the selected stars. The vertical slice contains 8,834 stars, and the horizontal slice contains 3,952 stars with 238 stars overlapping. We choose the vertical slice of
action space as $6 \lesssim \sqrt{J_R}\lesssim 14$~kpc\,km\,s$^{-1/2}$, $1655 \lesssim {L_z} \lesssim 1700$ kpc\,\kms\ as shown in Figure~\ref{fig:actionplotwstuff}. The $\sqrt{J_{R}}$ lower bound begins at the location of Ruprecht~147 since its presence indicates stars in that region are likely over 1~Gyr old, and implies that finding young stars there would be anomalous. This slice also captures part of a wrinkle structure (discussed in section \ref{sec:intro}). Our horizontal slice is of a similar thickness located at $11 \lesssim \sqrt{J_R} \lesssim 11.5$~kpc\,km\,s$^{-1/2}$, $1300 \lesssim {L_z} \lesssim 2050$~kpc\,\kms. This slice also captures a part of two wrinkles. 

\section{Methods} \label{sec:methods}

\subsection{Measuring \prot}\label{sec:measuring}
\begin{figure*}
\centering
\includegraphics[width=\textwidth]{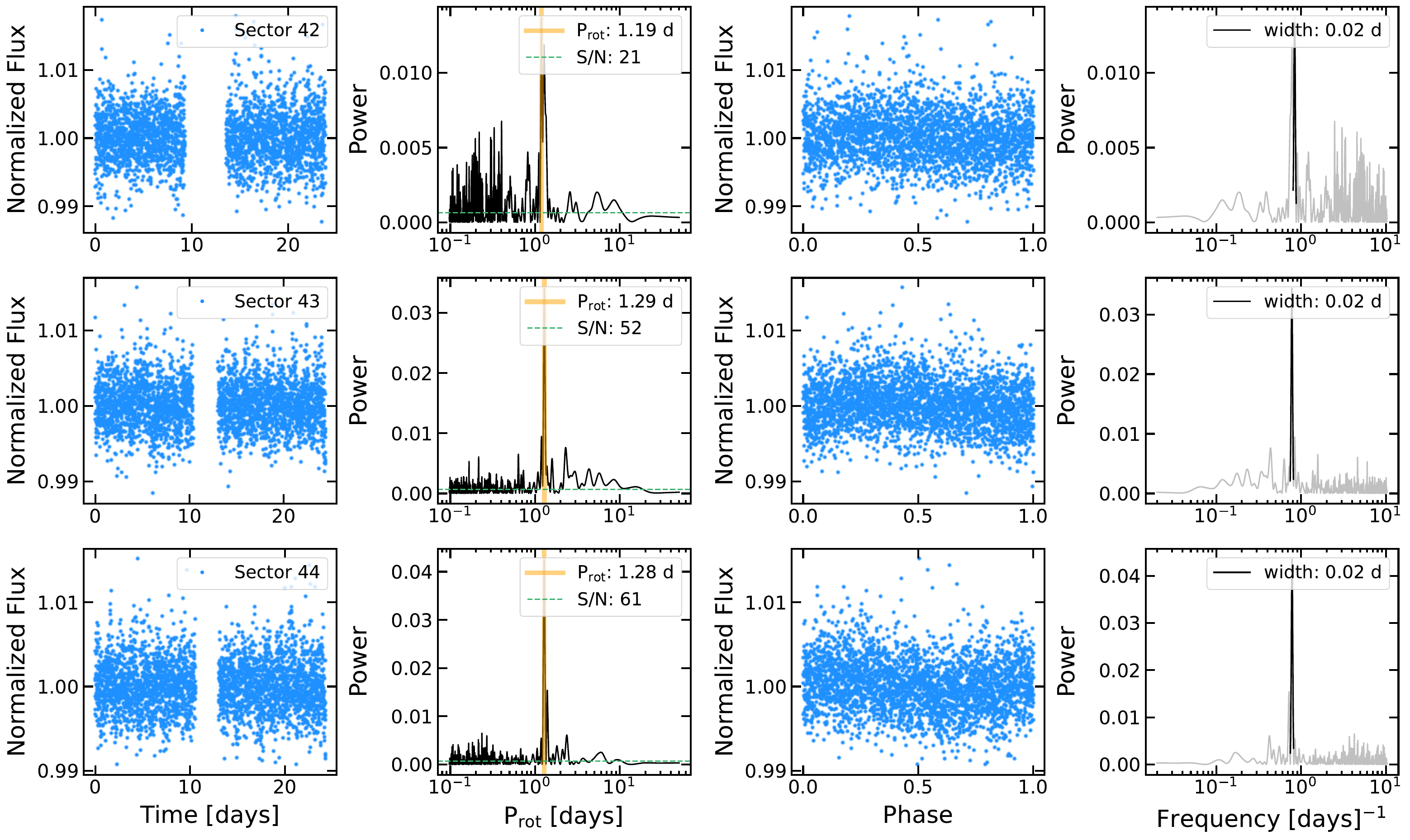}
\caption{Example of measurement and confirmation process for a star observed in sectors 42 (top row), 43 (middle row), and 44 (bottom row) that is detected in Pipelines I, II, and III. \textit{Left}: Star's light curve for sector. \textit{Middle-Left}: GLS periodogram with measured \prot\ associated with highest peak indicated in yellow, median power indicated in green, and calculated S/N of the highest peak. The S/N of the three sectors (21, 52, 61) spans our detection threshold of $40$. 
\textit{Middle-Right}: Phase-folded light curve with measured \prot. \textit{Right}: Periodogram in frequency space with primary peak fitted for width measurement (in black). This star has a strong S/N 
and small peak-width in two of three sectors, which means its \prot\ is a Pipeline I detection that is automatically confirmed as part of Pipeline II and considered a Pipeline II/III detection.}
\label{fig:snrexs}
\end{figure*}

The Lomb--Scargle (LS) periodogram is perhaps the most used form of \prot\ measurement given its efficiency and ability to use unevenly sampled data \citep{Lomb76,Scargle82,press1989,vanderplas18}. After defining a grid of periods or frequencies, sine curves with the defined periods/frequencies are fit to the signal. The result is an estimate of the Fourier power as a function of period/frequency. The second panels of Figure \ref{fig:snrexs} show the periodograms for an example star. Visually, we can see the \prot\ is $\approx 1.2$ days (left panels), and this is where the periodograms have their highest peaks. This is confirmed by phase-folding the light curve with the measured \prot\ as shown in the third panels. 


The Generalized-Lomb-Scargle (GLS) periodogram \citep{gls} is an iteration of the LS periodogram in which the sine wave fit is more rigorous and can account for measurement errors and offsets resulting in more precise \prot\ measurements. It has been used successfully with TESS data already \citep{Francys2022}. We use pyastronomy’s implementation \citep{pya}.

There should be at least 1.5--2 rotation cycles observed to reliably measure a \prot. For TESS's typical observational baseline of 27 days, this means we cannot trust measured \prot\ beyond 13.5--18 days. We define a periodogram with periods ranging from 0.1 to 50 days with a grid spacing of 0.001 days. If \prot\ $\geq 19$ days is measured, we expect this is likely a spurious signal, and our pipeline then chooses a new, shorter \prot\ corresponding to the next largest peak. This ultimately limits any measured \prot\ to $\leq 19$ days, but we leave the periodogram limit at 50 days for better constraints on our periodograms' peak-width measurements (introduced in section \ref{sec:pipelineI}).

For our analysis in section \ref{sec:results} and our injection and recovery test discussed in Appendix \ref{sec:injection}, we set the reliable \prot\ detection limit at 13 days. This limit matches other TESS rotation pipelines and guarantees two rotation cycles in a single TESS sector. Previous attempts at automating measuring rotation using single sectors of TESS data have only been successful for \prot\ $\leq 13$ days (\citealt{CantoMartins20,Holcomb,Kounkel_2022}), with measurements of longer periods more likely being associated with TESS systematics.

\prot\ measurements may return half or double the true \prot. This can result from sampling, short baselines, stellar spot evolution, or an intrinsically double-peaked signal (e.g. \citealt{basri_double_2018}). If a substantial peak (60\% of primary peak) is seen at double the period, then we automatically choose this for our \prot\ measurement as the code may be inadvertently folding the signal in half. If a longer harmonic is found at the pipeline \prot\ detection limit $\geq 19$ days, we choose the shorter harmonic as the measured \prot. We also calculate the window function for each light curve following \cite{vanderplas18}. We discard any \prot\ and peak-width that overlaps with the \prot\ returned from the primary peak of the window function, which is typically between 14--17 days.


\begin{figure*}
\centering
\includegraphics[width=\textwidth]{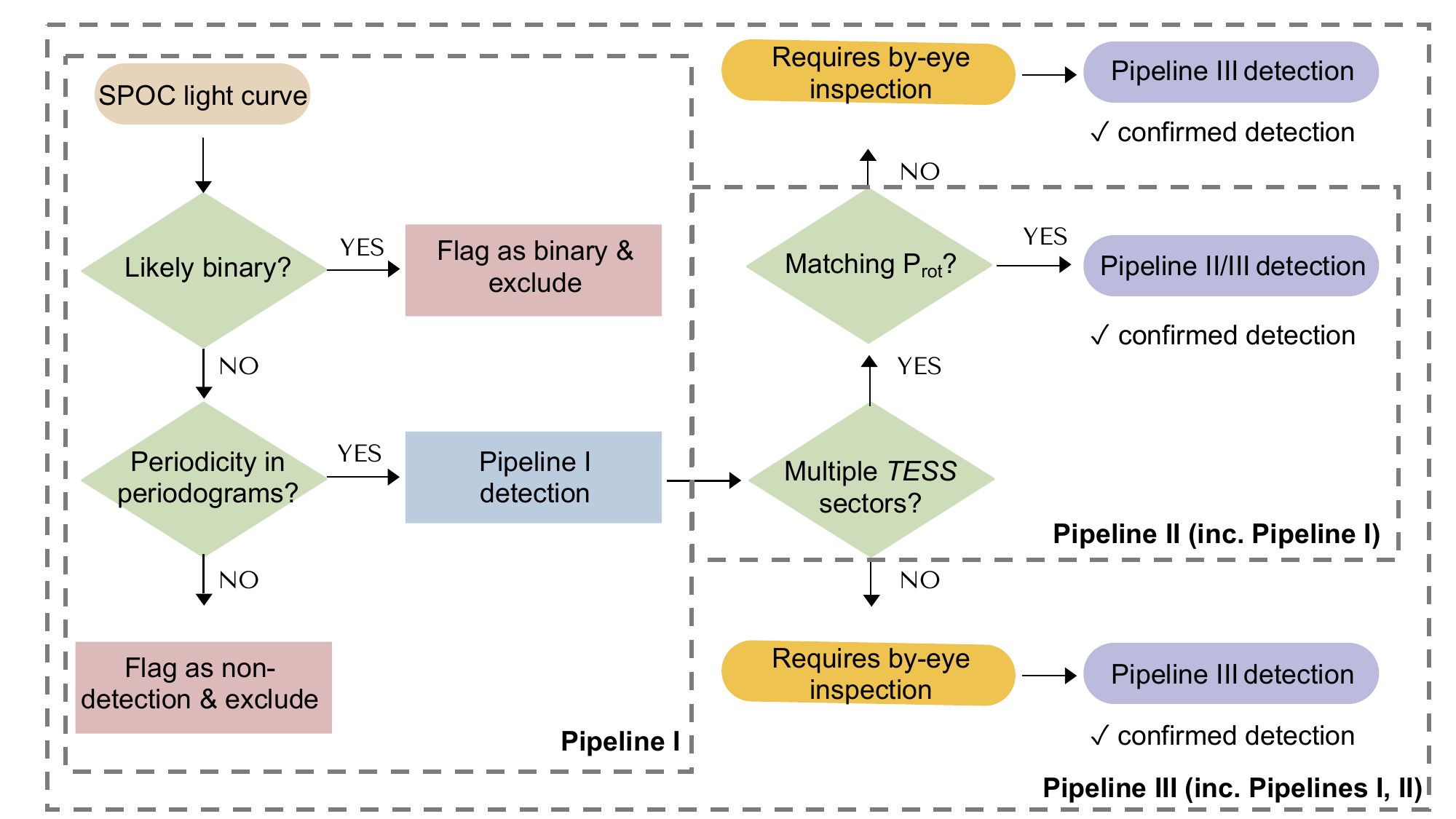}
\caption{Flowchart of the \prot\ pipelines. Starting with a star's light curve(s), we check if it is a candidate binary using ancillary data and discard if so. Next, we check for periodicity in the periodograms and identify stars with strong periodicity (Pipeline I detections). A Pipeline II detection is one in which a star has multiple sectors have Pipeline I detections, where the same \prot\ is measured. A pipeline III detection includes pipeline II detections plus any additional pipeline I detections that pass a by-eye inspection.}
\label{fig:flowchart}
\end{figure*}

\subsection{Confirming \prot}\label{sec:validation}

Instrument systematics and spot evolution require that a candidate \prot\ be subject to a confirmation process. For \prot\ in open clusters or samples of field stars, this has largely been done by eye since the number of stars is usually under a few thousand. To confirm larger samples like the entire Kepler or K2 fields, a series of metrics are used such as amplitude of the signal or primary peak height threshold (\citealt{mcquillan2014,reinhold20}). These thresholds are determined based on initial visual confirmation decisions and work best for data less subject to systematics than that from TESS (e.g. Kepler).

With TESS surveying the majority of the sky, it is necessary to develop an accurate confirmation that is at least partially automated, despite the systematics present. Automated machine learning tools such as random forest classifiers and convolutional neural networks have been used to measure \prot\ in Kepler (e.g. \citealt{rooster}) and TESS data. These methods do not necessarily outperform non-machine-learning methods for non-consecutive sectors of data at present \citep{Claytor22}.

We develop a quasi-automated confirmation process that significantly reduces the number of stars that require by-eye inspection. This process results in three pipelines that employ varying levels of confirmation. We illustrate these pipelines in Figure \ref{fig:flowchart} and outline the process below. Each subsequent pipeline includes the previous pipeline's steps, meaning that Pipeline III is the most comprehensive and thorough. Pipeline I and II are fully automated while Pipeline III adds visual inspection in select cases.

\subsubsection{Pipeline I: two metrics required to be considered a period detection, no by-eye inspection required}\label{sec:pipelineI}

 Before querying the star's light curve, we determine if the star is a candidate binary based on archival catalog information (see section \ref{sec:binaries} for how this is done). If it is, we exclude the star from further analysis. If single, we check if the star's light curve(s) displays significant periodicity by calculating two metrics for each periodogram. 
 
 First, we calculate the peak power over the median power, or signal-to-noise ratio (S/N) of the periodogram. We show examples of a star with strong periodicity in two of three sectors (S/N $>40$, Figure \ref{fig:snrexs}). 
 
Second, we measure the peak-width of the period detected as a diagnostic tool for \prot\ measurement quality within and across sectors. We use the width of the highest peak in the GLS periodogram (see right hand panels in Figure \ref{fig:snrexs}), which we calculate by fitting a Gaussian. We discard any stars that I) have a peak-width measurement $\geq 25\%$ of the candidate \prot\ or II) the peak-width measurement has failed. If the fitting fails, this indicates there was no peak-like feature to fit in the periodogram. This tends to happen at longer (and likely spurious) periods, where the periodogram feature is more plateau-like than peak-like.

If a star has at least one light curve with S/N $\geq 40$ (see determination of this value in Appendix \ref{sec:injection}) and a peak-width measurement $< 25\%$ of the candidate \prot, then we consider it a \textit{period detection} for Pipeline I and continue to the next pipeline. The final \prot\ reported for Pipeline I is the median \prot\ if there are Pipeline I detections in multiple sectors for a star.

\subsubsection{Pipeline II: multiple sectors with Pipeline I detections, no by-eye inspection required}
 
If the star passes Pipeline I with at least one detection, we check if the star has multiple sectors with Pipeline I detections \footnote{Given the systematics and data gaps in TESS, it is not yet possible to uniformly detrend light curves from multiple, non-consecutive sectors, stitch them together, and measure \prot\ longer than a single sector. We leave the maximum period in the periodogram at 50 days. However, inferring long \prot\ from multiple sectors of data is an area of active work (Hattori+ in prep, \citealt{claytor2023tess}).}. If there are multiple sectors, we see if the \prot\ from each sector are within 2x the peak-width including any 1/2$\times$ or 2$\times$ harmonics. This comparison is done by taking the median Pipeline I \prot\ detection and comparing each individual \prot\ measurement to the median. If there are an even number of detections, we choose the higher middle value as the median. If $\geq 2/3$ of the detections match the median within the uncertainties we measured in Pipeline I, the star's median \prot\ is considered a Pipeline II detection and reported as the final \prot. If we find a harmonic is present among sectors, we report the longer \prot\ as the final \prot. 

This form of automated confirmation is based on results from \cite{Rampalli21} and \cite{reinhold20}, which found that in spite of stellar spot evolution that occurs in stars observed with the K2 mission, it is expected that \prot\ can be recovered within 10--20\% across year-long gaps between observations excluding harmonic effects. Because of the shorter observational baseline and systematics, longer TESS \prot\ (more than 10 days) may differ by more than 20\% between sectors but still measure the same \prot. The peak-width measurement captures a similar level of uncertainty. 

\subsubsection{Pipeline III: by-eye inspection for stars that did not pass Pipeline II}

\begin{figure*}
    \centering
    \includegraphics[width=\textwidth]{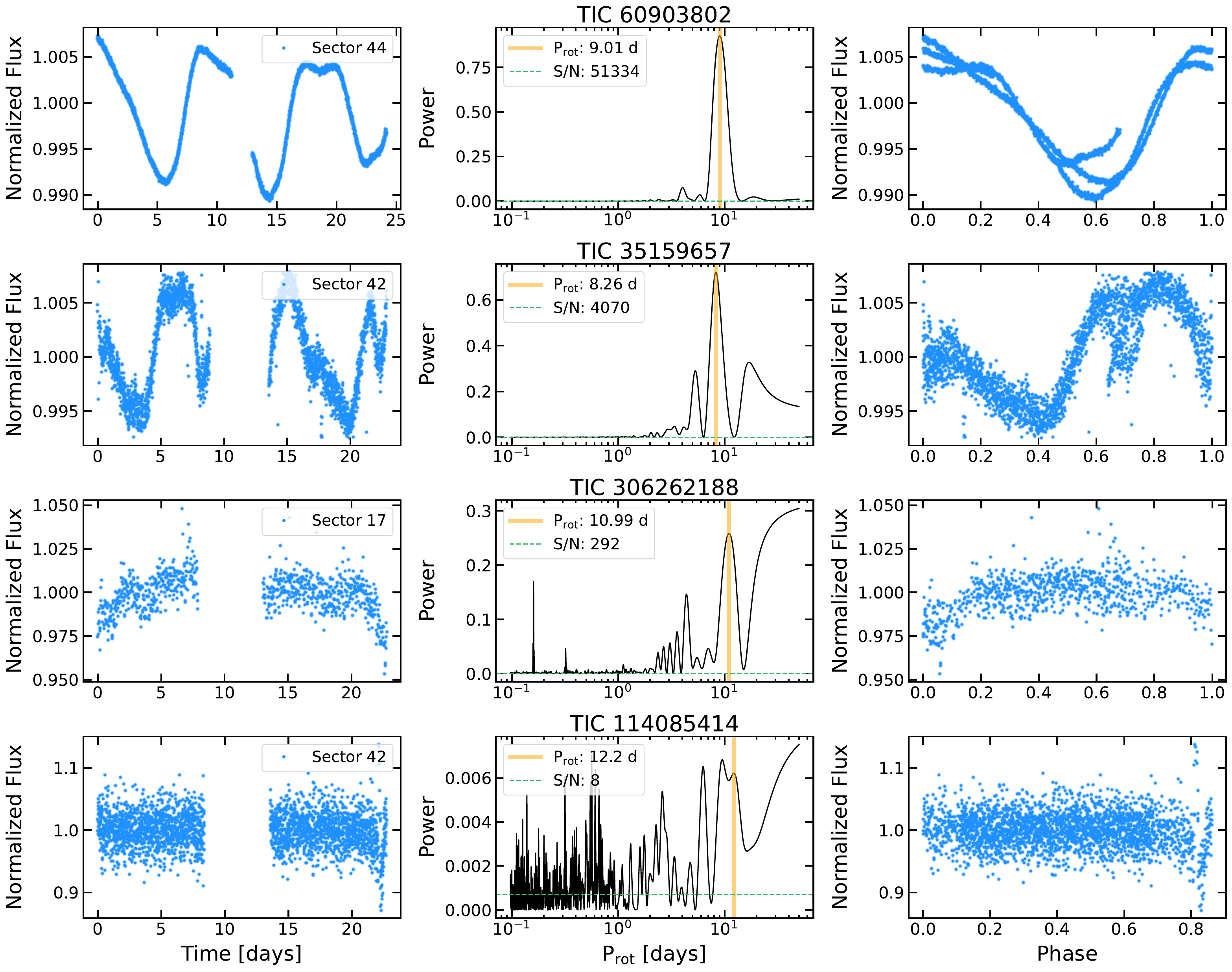}
    \caption{Examples of quality flag designations. Light curves are on the left, the corresponding periodogram in the middle, and phase-folded light curve on the right. The position of the detected period is indicated by the orange line. Top row: Q = 0. There is a clear 9 day periodic modulation in the light curve and a strong, sharp peak in the periodogram (literature \prot: 8.7 days). Second row: Q = 1. There is evidence of modulation measured at 8.3 days in the light curve, but it is weak due to instrument systematics and the data gap (literature \prot: 7.3 days). Third row: Q = 2. The light curve shows evidence for a long-term periodic trend likely due to systematics, which the periodogram chooses as the period instead of the masked and potentially astrophysically real period at 0.16 days (literature \prot: 0.32 days). Bottom row: Q = 3. Systematics completely dominate the light curve structure, and no astrophysical periodicity appears to be present (literature \prot: 0.43 days). }
    \label{fig:val}
\end{figure*}

A star with a period detection from Pipeline I might not be confirmed by Pipeline II for two reasons. The first case is stars where only a single sector has a Pipeline I period detection. The second case is where there are multiple sectors with Pipeline I period detections, but for which the periods do not agree. In these cases, the star is flagged for by-eye inspection in Pipeline III. To do a by-eye inspection for a star's candidate \prot, we check if the modulation in either the light curve of the phase-folded light curve matches the measured \prot. We use the quality check system (Q) used in \cite{Rampalli21}. We assign Q = 0 for obvious detections, Q = 1 for questionable detections (usually when there are obvious modulations in the light curve but an unclear period due to systematics), Q = 2 for spurious periodicity from instrument systematics, and Q = 3 for cases where the light curve is entirely dominated by systematics. We show examples of each classification in Figure \ref{fig:val}. Periods are considered visually confirmed for Q=0,1 and discarded for Q=3,4.

A star that passes either automated (i.e. Pipeline II detection) or by-eye inspection (i.e. Pipeline III detection) is considered a {\it confirmed detection}.

\subsection{Candidate Binary \& Subgiant Flagging}\label{sec:binaries}

Gyrochronology is intended to be applied to single stars. Tight binary systems' rotational evolution can be significantly altered due to each body exerting tidal forces on the other \citep[e.g.,][]{meibom2005, zahn2008}. This can lead to a spin up or down effect that deviates from the known single star age-rotation relations. A companion can also contaminate the rotation we are attempting to measure from the source star. Additionally, whether or not stars are tidally interacting or even physically associated, TESS pixels are large and can contain multiple stars. These effects lead to incorrect placement on the color-period plane and imprecise age inference, so we remove binaries from our sample. 

We also want to remove subgiants as while they might have solar colors, they were likely warmer when on the main sequence and beyond the Kraft break \citep{VanSaders13}. Thus, magnetic braking and gyrochronology may no longer apply despite the star having a rapid \prot.



We use the high-precision Gaia astrometry, GALAH spectroscopy, and TESS photometry in formulating the following tests to identify binaries:
\begin{enumerate}
\item Gaia DR3 includes a number of additional catalogs with non-single star solutions \citep{NSS}. This includes spectroscopic, photometric, and eclipsing binary observations. Solutions from the combinations of these observations are also provided in \citet{astrometric_bins} and \citet{eyer}. We check our stars are listed in any of these catalogs, flag them as binaries, and discard them from future analyses.

\item We examine the renormalized unit weight error (RUWE) measurement for each star. This is a goodness-of-fit measure of the single-star model fit to the source's astrometry. If a star has a RUWE~$> 1.4$, there is a strong likelihood the star has a companion (e.g., \citealt{ruwe2,ruwe1}). 

\item We check the GALAH DR3 catalog pre-main sequence star/binary flag column. Although pre-main sequence stars would be of interest given our goal of finding young stars (despite gyrochronology not being applicable), it is more likely for the stars to be binaries.

\item Each star in the TIC has a reported contamination ratio, or the ratio of flux from neighboring stars ($F_{\rm neighbor}$) to that of the source ($F_{\rm source}$) \citep{TIC}. We identify stars with contamination ratio (CR) $>20\%$ as visual binaries. This limit derives from the following procedure. Following \cite{Rampalli19}, the dilution ($d$) is equal to contribution of $F_{\rm source}$ to the total flux observed in the light curve ($F_{\rm total}$):
\begin{equation}
d =\frac{F_{\rm source}}{F_{\rm total}},
\end{equation}
 The contamination ratio (CR) is:
 \begin{equation}
 \text{CR} = \frac{F_{\rm neighbor}}{F_{\rm source}} = \frac{1-d}{d}.   \label{eqn:CRdil}
 \end{equation}

For a star with CR$=$20\%, we find using equation \ref{eqn:CRdil} that $d$ is 83\%, or alternatively that the neighboring star contributes 17\% of the flux. When $F_{\rm source}$ is constant and $F_{\rm neighbor}$ is varying by 5\% and contributes 17\% of the flux, the total flux variation is $0.83\times0 + 0.17\times0.05 = 0.0084$.


A flux variation of around 0.84\% 
has a 66\% chance of detection with our pipelines (see bottom right panels in Figure \ref{fig:exsector} in Appendix \ref{sec:injection}). 


\item Subgiants and equal-mass binaries can also pose as single-star interlopers. They are overluminous and sit above the single-star main sequence in a color-magnitude diagram. Following \cite{douglas2019,Rampalli21}, we fit a 6th order polynomial to the Gaia color, $G_{\rm BP}-G_{\rm RP}$ and absolute G-magnitude, $M_{g}$, color-magnitude diagram for the single-star rotator sequence in Praesepe. Stars that lay $> 0.4$ magnitudes above this fit are flagged as binary or evolved stars. 



\end{enumerate}

Failing any one of these tests is sufficient to be labeled a candidate binary or subgiant.

While we do apply these binary and subgiant cuts to stars with calculated actions, we do not apply these cuts to any of the clusters and MEarth groups when testing our pipelines except the contamination ratio cut. This maximizes the number of stars on which we can validate our pipelines unless a star was unresolved in TESS. Whether or not a star is a binary or subgiant does not affect period detection. Rather, it is the \prot\ interpretation and subsequent gyrochronological age that is affected. 

\subsection{Training, Validation, \& Testing with Literature \prot}

In order to understand how well our \prot\ pipelines work, we divide the five literature \prot\ catalogs into training, validation, and testing data sets. We begin with the training sample and use the validation sample to initially test the pipelines. Based on the validation sample results, we go back and adjust on the training sample to improve performance. The testing data set is used post-optimization to report final numbers (no changes can be made to the pipeline after reviewing the results for the testing data). 
We have broken up the catalogs in the following manner:
\begin{itemize}
\item Training: Hyades, Pleiades, 60\% MEarth
\item Validation: 50\% Praesepe, 50\% Psc-Eri, 20\% of MEarth
\item Testing: 50\% Praesepe, 50\% Psc-Eri, 20\% of MEarth
\end{itemize}

\begin{deluxetable*}{ccccccccc}
\centering 
\tabletypesize{\footnotesize} 
\tablecolumns{4}
\tablewidth{0pt}
\tablecaption{\prot\ pipeline versus literature \prot\ catalogs results}
\label{tbl:clusters_prot_ap}
\tablehead{
\colhead{Group} & 
\colhead{Pipeline} &
\colhead{\# of stars} &
\colhead{\% Recovery} & 
\colhead{\% Detection Recovery} &
\colhead {\% True Pos.} &
\colhead{\% True Neg.} & 
\colhead{\% False Pos.} & 
\colhead{\% False Neg.}
}
\startdata
 & I & 491 & 92 & 79 & 83 & \nodata & \nodata & 17 \\
Pleiades & II & 9 & 92 & 90 & 90 & \nodata & \nodata & 10 \\
 & III & 491 & 92 & 79 & 80 & \nodata & \nodata & 20 \\
\\
 & I &  46 & 93 & 93 & 100 & \nodata & \nodata & 0 \\
Pisces-Eridanus & II &  40 & 93 & 90 & 95 & \nodata & \nodata & 5 \\
 & III &  46 & 93 & 93 & 98 & \nodata & \nodata & 2 \\
\\
 & I & 208 & 90 & 83 &  91 & 89 & 11 & 9 \\
Praesepe & II & 118 & 90 & 84 &  92 & 93 & 7 & 8 \\
 & III & 208 & 90 & 82 &  90 & 94 & 6 & 10 \\
\\
 & I & 160 & 94 & 91 & 97 & 77 & 23 & 3 \\
Hyades & II & 107 & 94 & 92 & 96 & 79 & 21 & 4 \\
 & III & 160 & 94 & 90 & 93 & 89 & 11 & 7 \\
\\
 & I & 322 & 98 & 98 & 100 & 76 & 24 & 0 \\
MEarth & II & 104 & 98 & 98 & 99 & 87 & 13 & 1 \\
 & III & 322 & 98 & 97 & 97 & 95 & 5 & 3\\
\enddata
\tablecomments{Percentage recovered indicates whether 1, 2x, or 1/2x, the literature \prot\ was recovered with each pipeline within 2x the peak-width and within the reliable \prot\ detection limit ($\leq 13$ days). Percentage of detections recovered indicates whether 1, 2x, or 1/2x the literature \prot\ was recovered with each pipeline for a detection within 2x the peak-width and within the reliable \prot\ detection limit ($\leq 13$ days). True positive indicates a detection within the reliable \prot\ detection limit. True negative indicates a non-detection outside of the the reliable \prot\ detection limit. False positive indicates a detection outside of the the reliable \prot\ detection limit. False negative indicates a non-detection within the reliable \prot\ detection limit. The Pleiades and Pisces--Eridanus do not have true negative and false positive columns since $\leq 2$ stars have \prot\ $>13$ days. Pipeline II only considers stars with multiple sectors.}
\end{deluxetable*}

\begin{figure*}
\centering
\includegraphics[width=\textwidth]{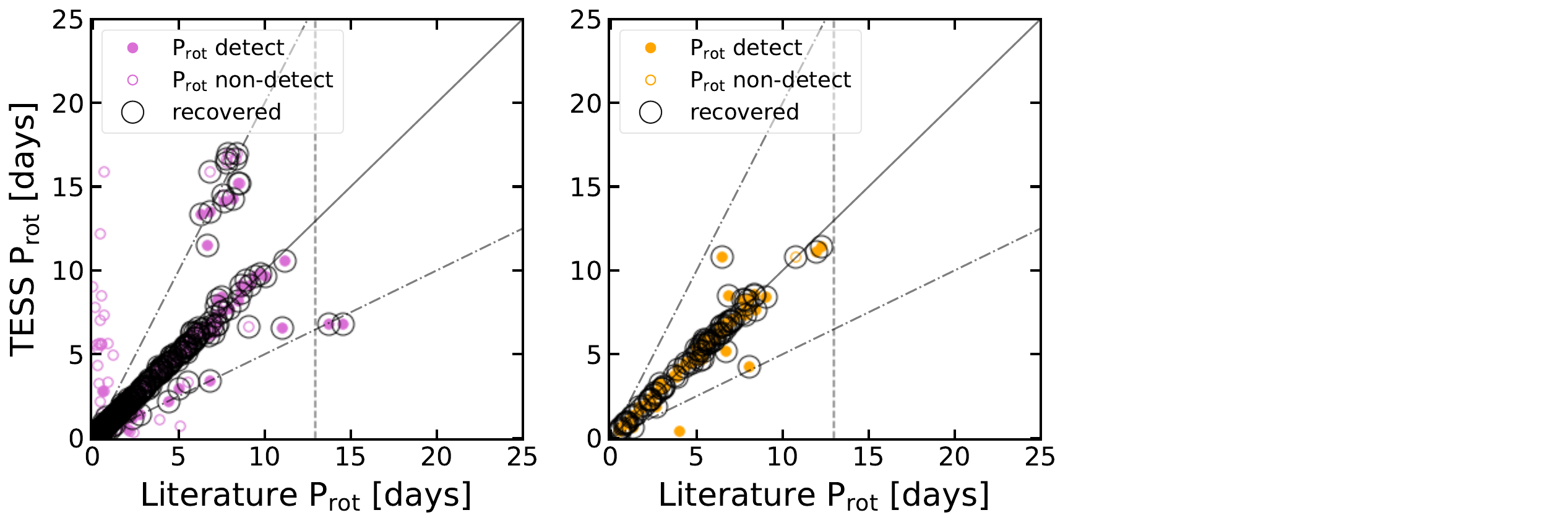}
\includegraphics[width=\textwidth]{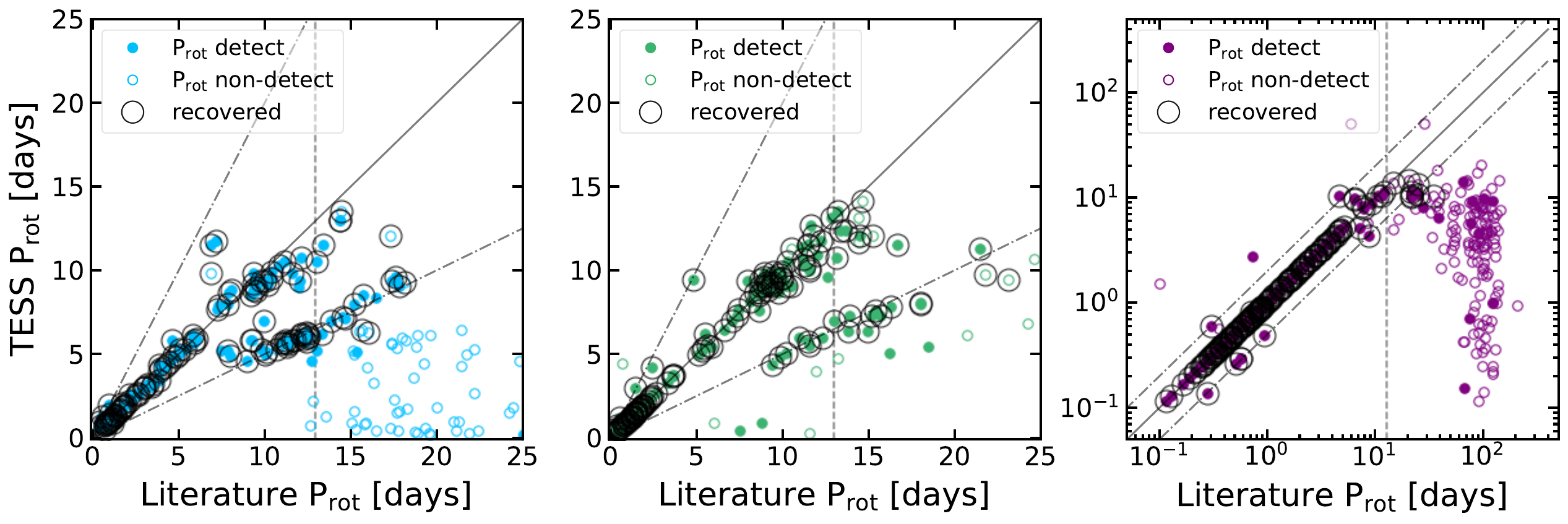}

\caption{Literature \prot\ versus TESS \prot\ for Pipeline III for Pleiades (pink), Pisces--Eridanus (orange), Praesepe (blue), Hyades (green), and MEarth (purple). Pipeline III detections considered recovered noted in black circles, and detection limit noted at 13 days with vertical dashed line.} 
\label{fig:litprot}
\end{figure*}

\begin{figure*}
\centering
\includegraphics[width=\textwidth]{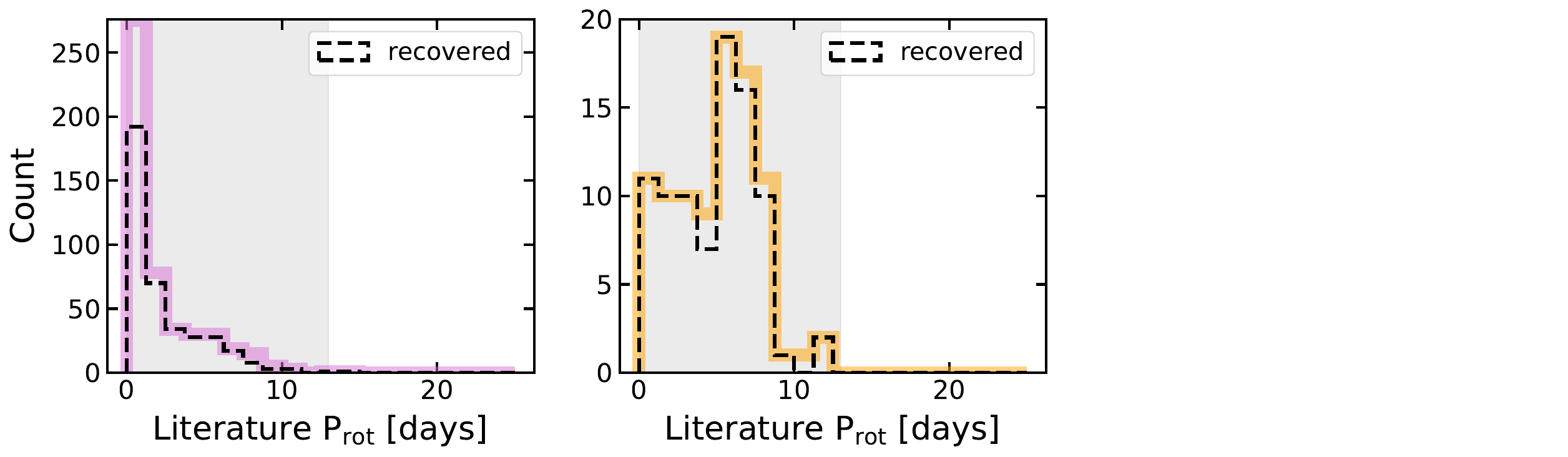}
\includegraphics[width=\textwidth]{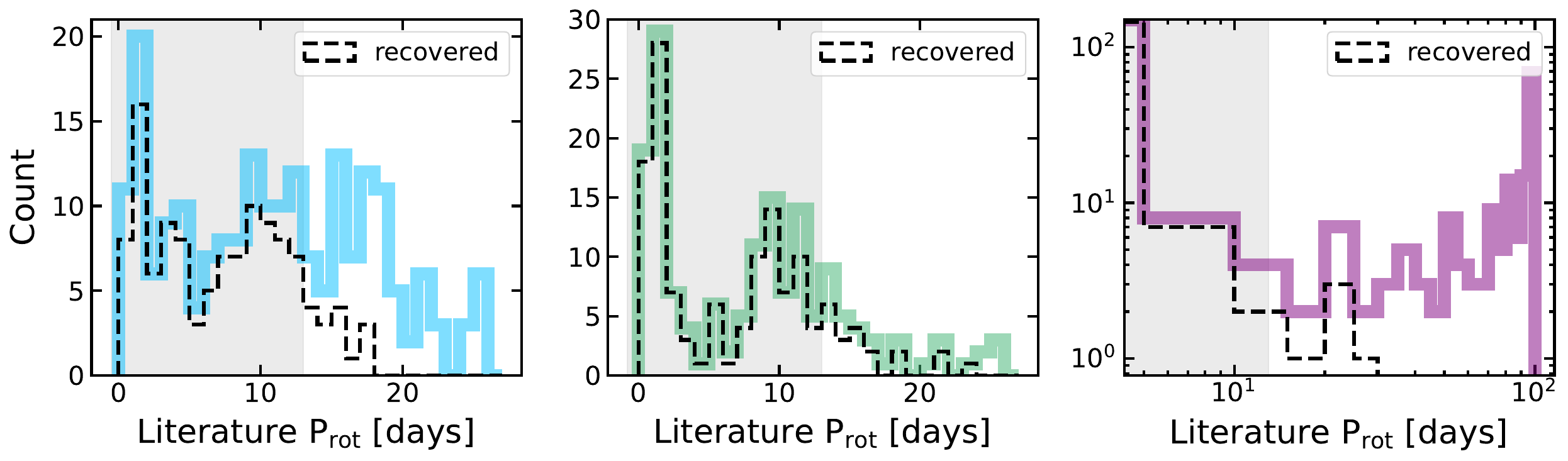}
\caption{Literature \prot\ distributions for Pipeline III detections for Pleiades (pink), Pisces--Eridanus (orange), Praesepe (blue), Hyades (green), and MEarth (purple). Recovered \prot\ noted by dashed black line, and gray regions indicate what we expect to detect with TESS ($\leq$ 13 days) shaded in gray.}
\label{fig:cmhists}
\end{figure*}

\begin{figure*}
\centering
\includegraphics[width=\textwidth]{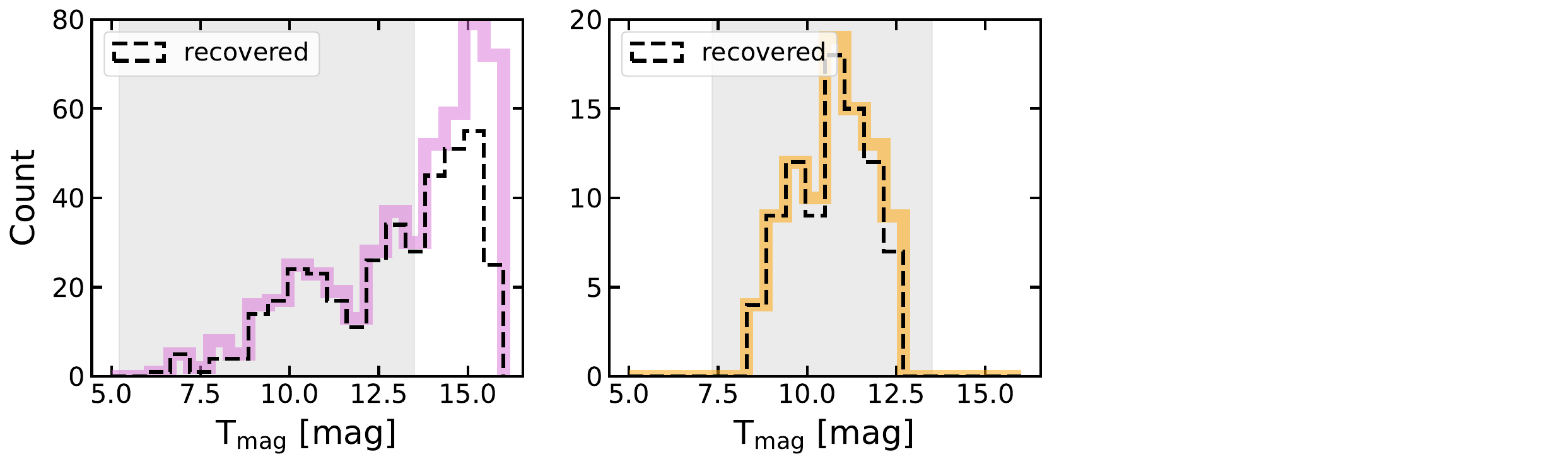}
\includegraphics[width=\textwidth]{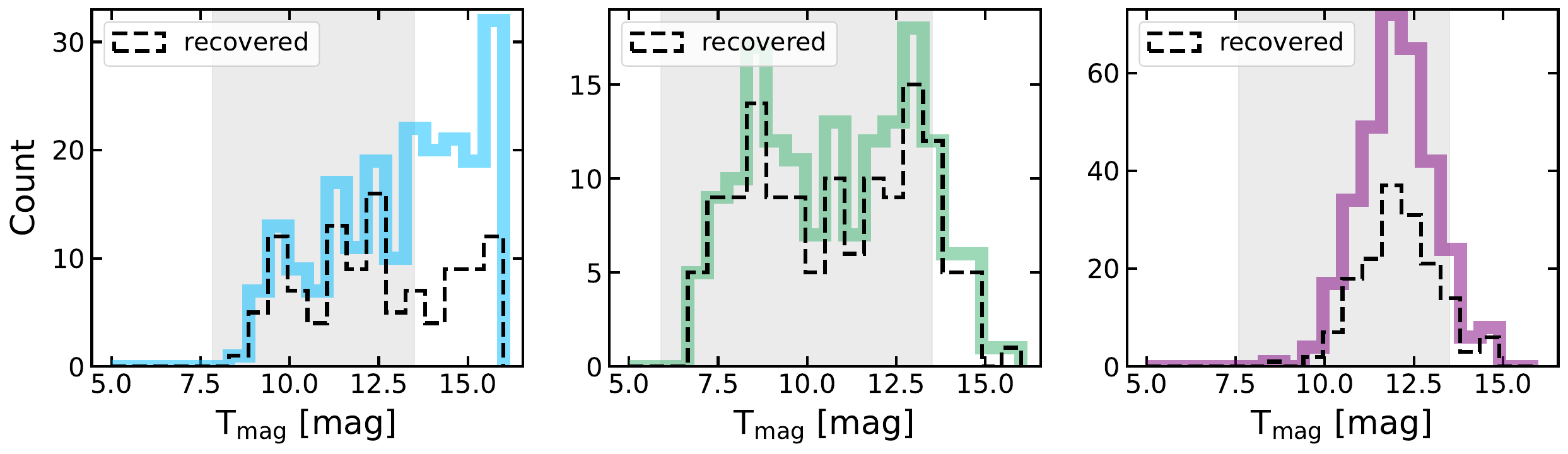}
\caption{TESS magnitude distributions for Pipeline III detections for Pleiades (pink), Pisces--Eridanus (orange), Praesepe (blue), Hyades (green), and MEarth (purple). Recovered stars noted by dashed black line, and gray regions indicate what we expect to detect with TESS-SPOC ($\leq$ 13.5 mag) shaded in gray.}
\label{fig:cmhists2}
\end{figure*}

\section{Pipeline Results\label{sec:results}}
 We calculate six metrics for each sample for each pipeline's definition of a detection to assess pipeline performance. In our calculations of completeness and recovery, we use \prot$ = 13$ days as the detection limit but do not impose this cut on our results, which occasionally include detections $ \geq 13$ days.

\begin{itemize}
\item Recovery rate for all stars: How many of our TESS \prot\ measurements are within 2x the measured peak-width of the literature \prot, considering only stars with literature \prot$< 13$ days. We also consider a \prot\ recovered if the \prot\ is within 2x the measured peak-width of a harmonic (1/2, 2) of the reported literature. A \prot\ does not have to be considered a detection by our pipeline for the correct literature \prot\ to be accurately measured.

\item Recovery rate for pipeline \textit{detections}: How many of our TESS stars had \prot\ \textit{detections} for the particular pipeline within 2x the measured peak-width of the literature \prot\ given the detection limit of 13 days? We also consider a \prot\ recovered if the \prot\ is within 2x the measured peak-width of a harmonic (1/2, 2) of the reported literature. 

\item True positive rate: How many of our TESS stars had \prot\ detections when a detection was expected (i.e., the true \prot\ was within the reliable \prot\ detection limit for TESS, $< 13$ days)? The detected \prot\ does not have to match the true \prot.

\item True negative rate: How many of our TESS stars had \prot\ non-detections when a non-detection was indeed expected (i.e. true \prot\ was outside the reliable \prot\ detection limit for TESS, $> 13$ days)?

\item False positive rate: How many of our TESS stars had \prot\ detections when a detection was not expected (i.e. true \prot\ was outside the reliable \prot\ detection limit for TESS, $> 13$ days)?

\item False negative rate: How many of our TESS stars had \prot\ non-detections when a detection was expected (i.e., the true \prot\ was within the reliable \prot\ detection limit for TESS, $< 13 $ days)?

\end{itemize}

\begin{figure*}
\centering
\includegraphics[width=0.6\textwidth]{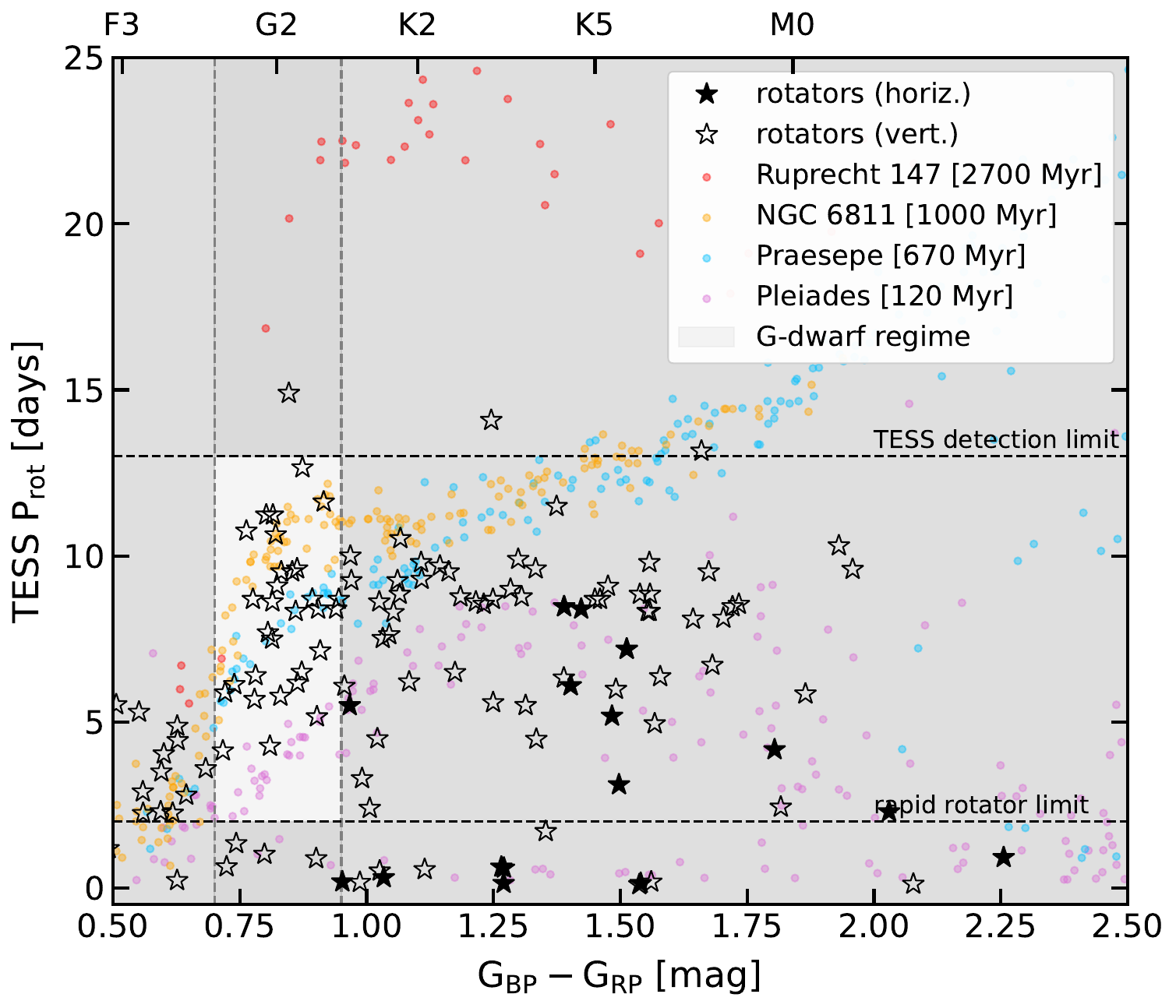}
\caption{Color--\prot\ distribution of likely single rotators for horizontal section of action space (black, filled stars) and vertical section of action space (unfilled stars) with benchmark clusters plotted for reference (120-Myr-old Pleiades in pink, 670-Myr-old Praesepe in blue, 1-Gyr-old NGC 68111 in yellow, 2.7-Gyr-old Ruprecht 147 in red). The most robust region for gyrochronology (G~dwarfs with reliable \prot\ measurements) is highlighted in the white region.}
\label{fig:wrinklecpd}
\end{figure*}

We report these values in Table \ref{tbl:clusters_prot_ap} for all five groups in the three pipelines. Pipeline II results have fewer stars than the other two pipelines since it only considers stars with multiple sectors of data as shown in Figure \ref{fig:flowchart}. While we have divided these groups up for the training and validation phases, we combine the datasets used for training and validation when reporting numbers since they were all subjected to the same analyses. We repeat this analysis with the test set: 50\% of the Pisces--Eridanus, 50\% of the Praesepe, and 20\% of the MEarth samples. The results are shown in Table 
\ref{tbl:clusters_prot_ap_test} in Appendix \ref{sec:testset}. The numbers are very similar to those in Table \ref{tbl:clusters_prot_ap} with an average recovery rate of 92\%, true positive rate of 92\%, true negative rate of 89\%.

For all five samples, the overall recovery rate is $\geq 90\%$. When considering recovery of each pipeline's \textit{detections}, the rates generally remain similar except for the Pleiades and Praesepe, whose distances of 136 and 186 pc lead to magnitude limitations for TESS light curves (discussed further below). 

True positive rates across groups and pipelines never fall below 80\%. The lowest true negative rate is 76\% for MEarth after running Pipeline I; this is not unexpected since Pipeline I has the least number of confirmation tests. The true negative rate increases to 87\% and then 95\% for MEarth with Pipelines II and III and thus subsequent confirmation checks. The Hyades has a similar true negative rate of 77\%. The higher false positive rate is because many Hyads have \prot\ just slightly $> 13$ days. As can be seen in Figure \ref{fig:litprot}, we identify a period for many of these stars at $<13$ days which are then counted as false positives given our imposed detection limit of 13 days.

With Pipeline I, true positive rates are strong ($83$ to $100\%$). However, when looking at the associated recovery rates for detections, or how well Pipeline I does measuring the \textit{correct} \prot\ for detections, we see that the accuracy is somewhat lower ($79$ to $98\%$). The true negative rates are quite strong as well with the exception of MEarth and the Hyades sample as discussed previously. 

Pipeline II only considers stars with multiple sectors of data as a means of confirming \prot. This additional confirmation test generally increases true negative rates, indicating that it is successful in identifying bad detections.

Pipeline III, our most rigorous pipeline, includes by-eye inspection. For the Pleiades, Pisces--Eridanus, Praesepe, Hyades, and MEarth, by-eye inspection of stars with mismatching \prot\ in multiple sectors occurred for 0\%, 6\%, 10\%, 16\%, and 11\% of the stars in the five respective groups. By-eye inspection is also invoked for single sectors, which occurred for 98\%, 7\%, 7\%, 13\%, and 42\% of the samples respectively. 

For Pipeline III, recovery rate for detections and true positive rates decrease slightly across all groups compared to Pipeline I. Most \prot\ detections in Pipeline I are labeled true positives despite potentially having inaccurately measured \prot\ because there are no additional confirmation steps in this pipeline; any star's light curve with a significant peak in a periodogram is deemed a detection. Adding the by-eye inspection and/or comparison of \prot\ across sectors decreases false positives; the trade-off is that it can move stars originally classified as true positives in Pipeline I to false negatives in Pipelines II and III. True negative rates increase from Pipeline I, especially for MEarth and the Hyades, indicating the importance of by-eye inspection. 

The overall trend going from Pipeline I, to II, to III is for improved true negative rates with only modest decreases in true positive rates. This is strongest in the MEarth sample. Given our motivations to explore ages of stars in high eccentricity orbits, (e.g. mostly old, field stars with some young interlopers), the MEarth sample is our closest proxy. Thus, going forward, we will choose Pipeline II or Pipeline III, which best improve chances of eliminating false positives.

We further investigate the reasons behind the statistics of our pipeline. We show our literature versus TESS \prot\ plots for the five groups in Figure \ref{fig:litprot} for Pipeline III. In comparison to the detections found at $< 13$ days, very few detections are found for stars with longer periods (these are primarily in Praesepe and MEarth). In the case of Praesepe and the Hyades, we measure the half-period harmonics for a fair number of the longer rotators.

In Figure \ref{fig:cmhists}, we project these results into literature \prot\ space. These extend to the length of a TESS sector, 27 days, to illustrate where our pipeline starts failing. For MEarth, we extend the histogram out to 100 days in log space. We highlight in grey the reliable \prot\ detection range. We show the TESS magnitude distributions in Figure \ref{fig:cmhists2}, highlighting in grey the TESS-SPOC magnitude regime. The targets for our period search --- young G~dwarfs --- are generally expected to fall within the grey regions. The near-perfect recovery of Pisces--Eridanus is expected since both use TESS data. However, the method of light curve extraction and confirmation by \citeauthor{PisEri} differs from ours yet we measure the same \prot. This validates the automated aspects of our pipeline since all of the \prot\ measured in \cite{PisEri} were visually inspected, in contrast to our pipelines.

Figure \ref{fig:cmhists2} demonstrates that the lower period recovery rates for detections in Pleiades and Praesepe is due to their distances. Because Praesepe has rotators beyond the 13 day limit, most of which are also faint, this further contributes in failure to recover \prot. Less than 60\% of \prot\ are recovered beyond 12 days (Figure \ref{fig:cmhists}). The rotators beyond 13 days are mostly non-detections despite the true \prot\ being recovered in some cases (Figure \ref{fig:litprot}). This is in contrast to the Hyades, which is close by. In the Hyades, rotators beyond 13 days are mostly detected with the true \prot\ recovered. Recovery rate decreases by 29\% from 11--13 days, with the sharpest drop in recovery ($\leq 60\%$) at 15 days. The MEarth sample has a drop to $< 50\%$ recovery in the same 10--15 day range. Figures \ref{fig:cmhists} and \ref{fig:cmhists2} illustrate the motivation for our 13 day limit (see also Appendix \ref{sec:injection}).

\section{Testing Pipeline II on stars in high eccentricity orbits}
\label{sec:actionresults}

\subsection{Pipeline Performance}
Given the promising results from our pipelines, we remove binaries and query TESS light curves for the vertical (8,834 stars) and horizontal slices (3,952 stars) of action space shown in Figure \ref{fig:actionplotwstuff}. After applying Pipeline II on the 2,887 and 1,198 likely single stars with TESS-SPOC PDC-SAP light curves for multiple sectors, we are left with 302 and 60 rotators respectively. The difference in number of rotators found in each slice is expected as there are more stars in the vertical slice. This region also samples relatively lower radial action space, where young stars are more likely to populate. 

We do an additional visual inspection following the quality flag guidelines outlined in Figure \ref{fig:val} and keep stars with Q=0,1. Post-inspection, we are left with 127 rotators out of 302 stars in the vertical slice, 17 rotators out of 60 stars in the horizontal slice, and three stars overlapping. This suggests a false positive rate of $\approx 60-70\%$ and is significantly higher than the false positive rates in the MEarth sample, which most resembles this sample of mostly field stars. We attribute this to the fact that 40\% of the stars in MEarth only had one sector of data and were therefore immediately flagged for by-eye inspection, increasing the true negative rates. While this implies that Pipeline II can be improved in the automatic vetting of multi-sector stars, Pipeline II has significantly reduced the number of stars normally needing by-eye inspection (4,085 to 359). 


\subsection{Rotator Distribution in Color Space}

We show the rotator distribution (filled stars are from the horizontal slice, unfilled from vertical) in color--\prot\ space in Figure \ref{fig:wrinklecpd} with the benchmark clusters' \prot\ for reference. We correct the Gaia colors for extinction, which is minimal given the proximity of the stars. We determine the $A_{v}$ from the calculated reddening in the Gaia bandpasses using Bayestar19 \citep{Green3D2019}. The rotation periods detected correspond to periods typical for Pleiades-age (120 Myr) single stars to beyond the age of NGC 6811 (1 Gyr). 

While we have removed binaries, there is likely still binary contamination in our sample. 
\prot\ work with Kepler and K2 have shown that rapid rotators with \prot\ $\leq 2$ days largely tend to be binary stars undergoing tidal interactions \citep{Douglas16,Simonian19}. Following \cite{Kounkel_2022}, we flag stars with $\prot < 2$ days, and we consider the most robust region for gyrochronology to be $2 < \prot\ < 13$ days. While there are young G~dwarfs (11-20\% for 87-120 Myr) that can have $< 2$ day periods and be misclassified as binaries, this makes up a very small portion of the entire sample \citep{Rebull18,Boyle22}. This is also on the same timescale of convergence onto the main sequence, where gyrochronology may not yet be applicable \citep{Bouma23}. Recall 13 days was chosen as the upper limit to account for TESS's short observational baseline, to minimize misclassifying its orbital period systematics as astrophysical periodicity, and based on injection and recovery tests (Appendix \ref{sec:injection}). These limits are shown as grey-dashed horizontal lines in Figure \ref{fig:wrinklecpd}. We demarcate the G-dwarf regime with grey-dashed vertical lines, where gyrochronology will work the best since age--rotation relations are best calibrated for G~dwarfs. There are 30 G~dwarfs within these period and color limits that have \prot\ Pipeline II detections. Stars outside of this period and temperature regime may still be useful for indicating youth. For example, K~dwarfs with \prot\ where spin-down has stalled still indicate youth, as do stars that have not yet converged to a gyrochronology track.

Though rapid rotation suggests youth, it is not a definitive indicator. We will obtain pre-existing or new spectra to provide an additional check for youth with abundance and/or activity measurements. One such elemental abundance is lithium, \li. Lithium is one of the most sensitive elements and burns at fairly low temperatures within a star. As a low-mass star ages, its measured \li\ ratio decreases due to the burning \citep{litimescales}. Using spectra from groups of coeval stars of known ages, empirical Li--age relations have been derived \citep{baffles,EAGLES} and have been used to confirm stellar ages in combination with gyrochronology (e.g. \citealt{benthyme,Groupx}).


Using our kinematics catalog, we do a preliminary investigation 
\li\ abundance measurements that may exist for rotators. 
Only 15/141 unique stars had GALAH spectra. Five of the fifteen had reported Li abundances, and all were $\li\ \leq -0.25$ dex indicating they are on the order of several Gyr old \citep{LiAbundances}. This strongly motivates the need for additional confirmations of youth and binary screening in our future work.

\begin{figure*}
\centering
\includegraphics[width=0.55\textwidth]{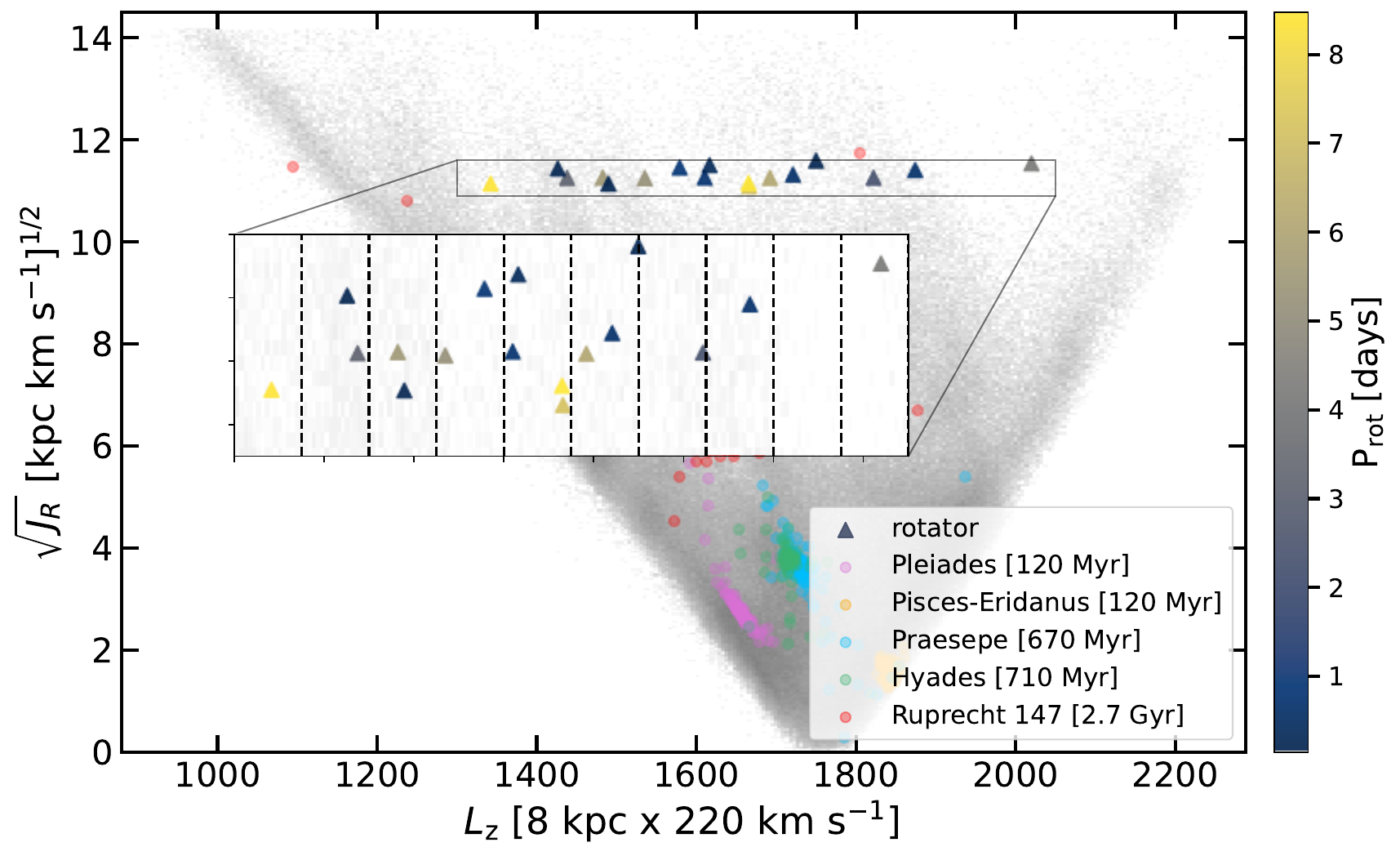}
\includegraphics[width=0.4\textwidth]{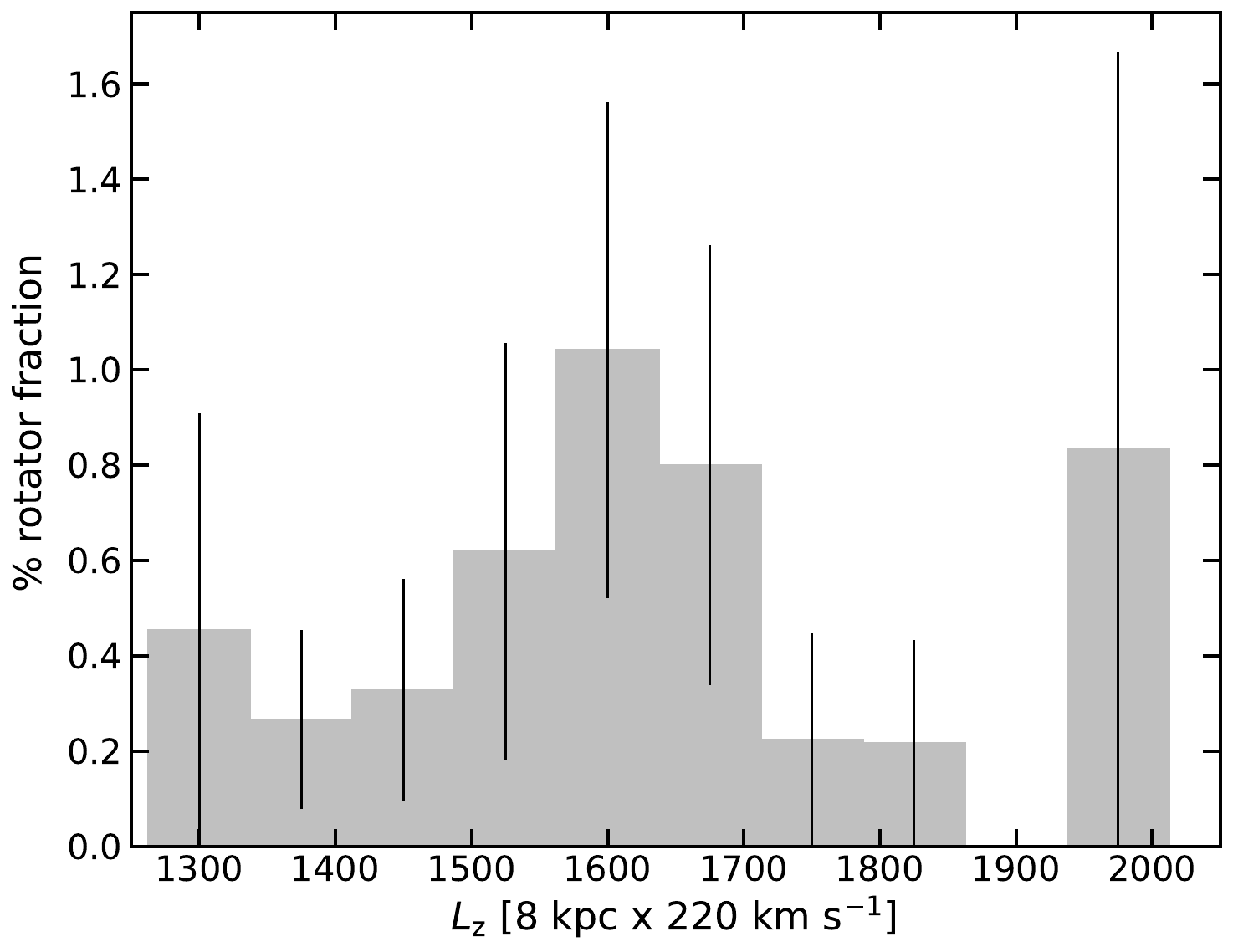}
\caption{\textit{Left:} Rotator distribution in horizontal slice of action space with benchmark clusters overplotted as shown in Figure \ref{fig:actionplotwstuff}. Bins from right-hand panel indicated by dashed vertical lines in inset. \textit{Right:} Fraction of rotators to all stars found in bins of angular momentum in z-direction.}
\label{fig:wrinkleplotprot_lz}
\end{figure*}

\begin{figure*}
\centering
\includegraphics[width=0.55\textwidth]{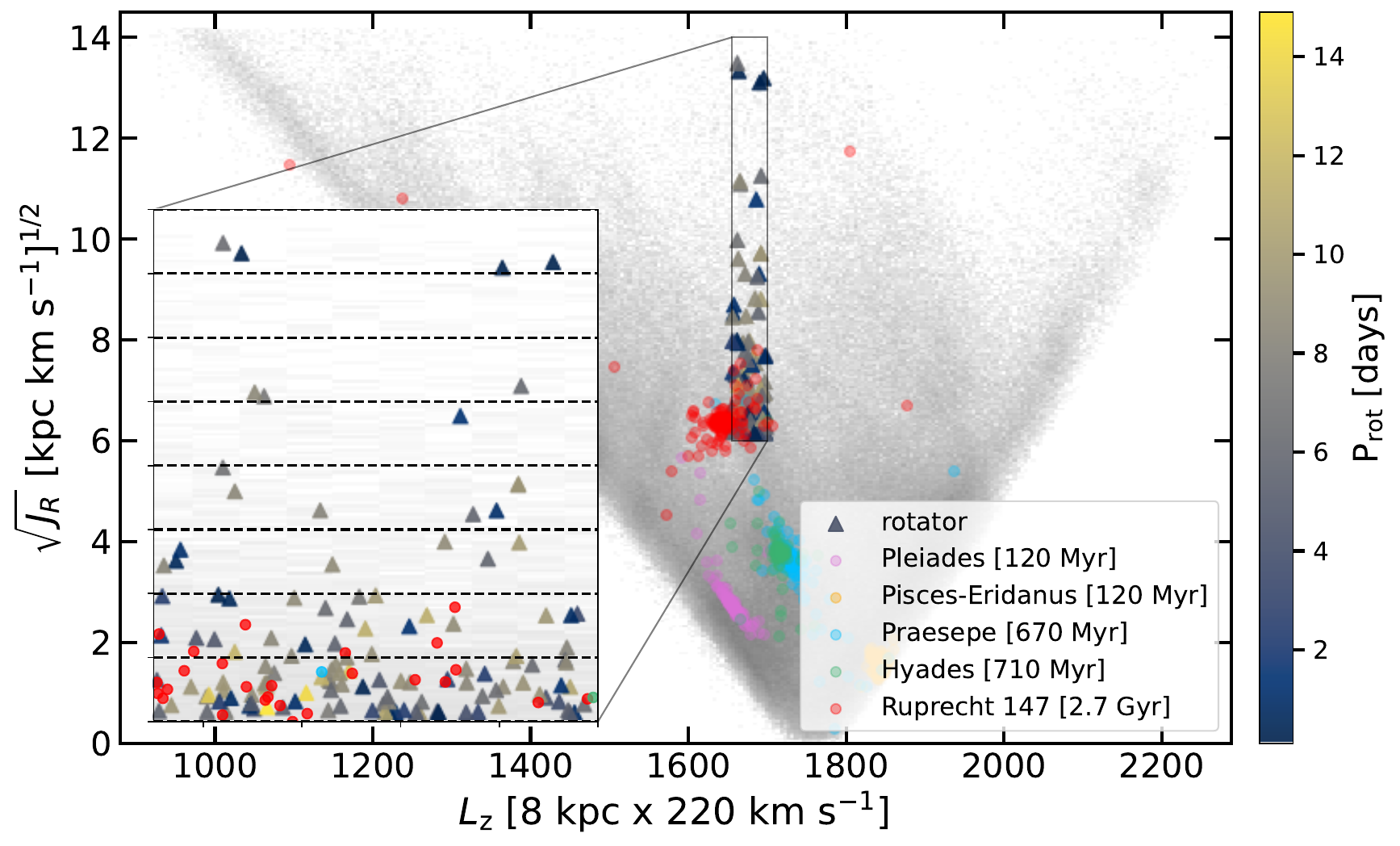}
\includegraphics[width=0.4\textwidth]{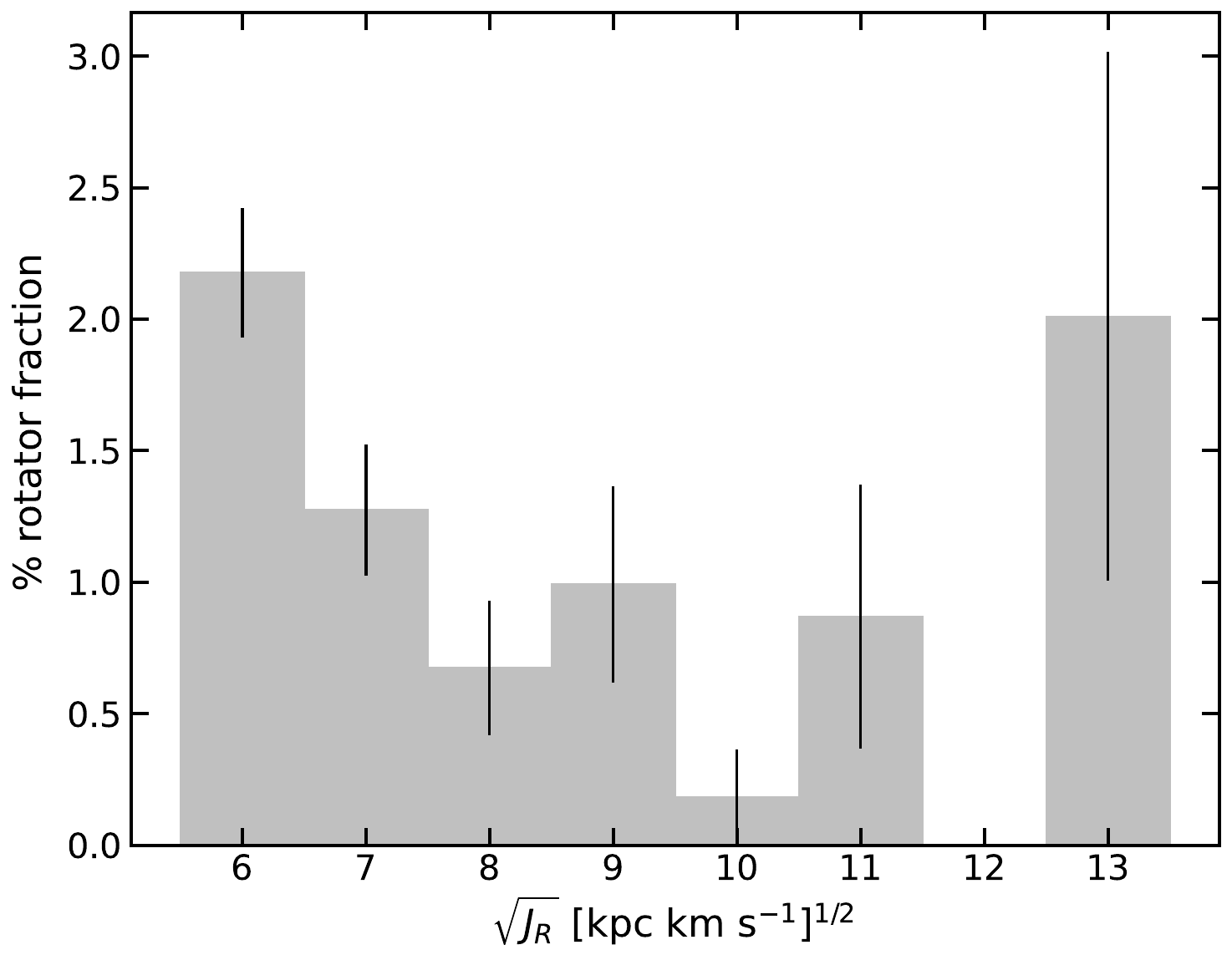}
\caption{\textit{Left:} Rotator distribution in vertical slice of action space with benchmark clusters overplotted as shown in Figure \ref{fig:actionplotwstuff}. Bins from right-hand panel indicated by dashed horizontal lines in inset. \textit{Right:} Fraction of rotators to all stars found in bins of square root of radial action.}
\label{fig:wrinkleplotprot_jr}
\end{figure*}

\subsection{Rotator Distribution in Action Space}
While some rotators in this sample will be identified as interloping old stars, we have found stars with rotation indicative of youth in highly eccentric orbits as indicated by their radial actions. We show the distribution of rotators in action space in the left hand panel of Figures \ref{fig:wrinkleplotprot_lz} and \ref{fig:wrinkleplotprot_jr} color-coded by the stars' \prot\ as well as the benchmark clusters plotted for reference. In the right hand panel of Figures \ref{fig:wrinkleplotprot_lz} and \ref{fig:wrinkleplotprot_jr}, we show the percent fraction of rotators found in binned values of radial action (vertical slice) and angular momentum (horizontal slice) with error bars. These bins are indicated in dashed lines in the left hand panel insets. We do not make the cuts on \prot\ or color delineated in Figures \ref{fig:wrinklecpd} as we are not inferring ages for these stars just yet and are interested in examining the full rotator distribution.  

In the horizontal slice the distribution of rotators is flat considering the errors. Visually, there is a peak at $L_{z} \sim 1600, \sqrt{J_{R}}\approx 11$, which corresponds to part of one of the many overdensities in action space that we call wrinkles (discussed in section \ref{sec:GAS}).

In the right hand panel of Figure \ref{fig:wrinkleplotprot_jr}, we see a significant peak at $\sqrt{J_{R}} \sim 6$. This is an exciting region dynamically, which we will explore in depth in following papers in this series. There are older associations like Ruprecht 147 (2.7 Gyr) whose stars could be in these orbits. This indicates some of the rotators we have identified could be from older associations such as these. However, we also note the presence of $> 10$ rotators that follow the Pleiades-like color-\prot\ distribution in Figure \ref{fig:wrinklecpd}, which is unexpected this high in radial action space. This general region also includes part of a wrinkle as well.

\section{Conclusions \& Future Directions}\label{sec:future}

The combination of the Gaia and TESS data offers an unprecedented opportunity to understand the dynamical history of the Milky Way using stellar ages with gyrochronology and kinematics. We build a series of TESS \prot\ measurement and confirmation pipelines that we test on 1,560 stars with literature \prot\ from 5 groups: the Pleiades, Pisces--Eridanus, Praesepe, the Hyades, and the MEarth sample. On average, we find a recovery rate of 88\%, a true positive rate of 91\%, and a true negative rate of 89\%. We find that our pipelines predominantly fail when stars are distant and thus faint like the cooler members of the Pleiades (136 pc) and Praesepe (186 pc). 

Given the promising \prot\ pipeline results (Table \ref{tbl:clusters_prot_ap}), we apply our Pipeline II to a vertical and horizontal slice in Galactic action space, which we calculate using kinematics from Gaia DR3 and radial velocities from GALAH and APOGEE when possible (Figure \ref{fig:actionplotwstuff}). Each slice contain 1--2 overdensities of stars in highly eccentric orbits \cite{Trick19} identified, which we call wrinkles. We find 141 unique rotators out of the 12,786 in these selected regions. Five stars have GALAH \li\ measurements that indicate they are interloping, several-Gyr-old stars (beyond the age where gyrochronology is applicable). When looking at the other rotators' Gaia color--\prot\ distributions, we see the stars' rotation periods suggest they are 120 Myr to 1 Gyr old (Figure \ref{fig:wrinklecpd}). What is especially interesting is the rotator distribution at $L_z \approx 1675$ and $\sqrt{J_R} \approx 6$. This could be an area where older $\approx 1$ Gyr associations live but perhaps also 120-Myr stars that could have been dynamically heated to these eccentric orbits.

The identification of rapid \prot\ $<13$ days for stars in highly eccentric orbits motivates our future studies. We will apply our rotation pipelines to all of the stars at highly eccentric orbits ($\sqrt{J_{R}} \geq 6$). Once \prot\ are measured, we will age-date these stars using gyrochronology relations and confirm or refute their youth with spectroscopic observations. The confirmation of young stars at high radial actions in turn will allow us to understand the dynamics that caused their orbits to become highly eccentric.

\vspace{17pt}
We thank the anonymous referee for constructive comments that helped
us improve the manuscript. We thank the THYME Collaboration, the Cool Stars 21 Conference, the Scialog conference, Graham Edwards, Alex Mule, Aylin Garcia Soto, Adrian Price-Whelan, and the Dartmouth graduate student community for helpful discussions. We thank the CCA for hosting KD's sabbatical and the group for in-person collaborative meetings. This work was funded by the NASA ADAP (21-ADAP21-0134). RR thanks the LSSTC Data Science Fellowship Program, which is funded by LSSTC, NSF Cybertraining Grant \#1829740, the Brinson Foundation, and the Moore Foundation; her participation in the program has benefited this work.

Funding for the TESS mission is provided by NASA’s Science Mission directorate.

This work has made use of data from the European Space Agency (ESA) mission Gaia (https://www.cosmos.esa.int/gaia), processed by the Gaia Data Processing and Analysis Consortium (DPAC, https://www.cosmos.esa.int/web/gaia/dpac/consortium). Funding for the DPAC has been provided by national institutions, in particular the institutions participating in the Gaia Multilateral Agreement.This work made use of the Third Data Release of the GALAH Survey (Buder et al. 2021). 

The GALAH Survey is based on data acquired through the Australian Astronomical Observatory under programs: A/2013B/13 (The GALAH pilot survey); A/2014A/25, A/2015A/19, A2017A/18 (The GALAH survey phase 1); A2018A/18 (Open clusters with HERMES); A2019A/1 (Hierarchical star formation in Ori OB1); A2019A/15 (The GALAH survey phase 2); A/2015B/19, A/2016A/22, A/2016B/10, A/2017B/16, A/2018B/15 (The HERMES-TESS program); and A/2015A/3, A/2015B/1, A/2015B/19, A/2016A/22, A/2016B/12, A/2017A/14 (The HERMES K2-follow-up program). 
We acknowledge the traditional owners of the land on which the AAT stands, the Gamilaraay people, and pay our respects to elders past and present. This paper includes data that has been provided by AAO Data Central (datacentral.org.au).

Funding for the Sloan Digital Sky Survey V has been provided by the Alfred P. Sloan Foundation, the Heising-Simons Foundation, the National Science Foundation, and the Participating Institutions. SDSS acknowledges support and resources from the Center for High-Performance Computing at the University of Utah. The SDSS web site is \url{www.sdss.org}. SDSS is managed by the Astrophysical Research Consortium for the Participating Institutions of the SDSS Collaboration, including the Carnegie Institution for Science, Chilean National Time Allocation Committee (CNTAC) ratified researchers, the Gotham Participation Group, Harvard University, Heidelberg University, The Johns Hopkins University, L’Ecole polytechnique federale de Lausanne (EPFL), Leibniz-Institut fur Astrophysik Potsdam (AIP), Max-Planck-Institut fur Astronomie (MPIA Heidelberg), Max-Planck-Institut fur Extraterrestrische Physik (MPE), Nanjing University, National Astronomical Observatories of China (NAOC), New Mexico State University, The Ohio State University, Pennsylvania State University, Smithsonian Astrophysical Observatory, Space Telescope Science Institute (STScI), the Stellar Astrophysics Participation Group, Universidad Nacional Autonoma de Mexico, University of Arizona, University of Colorado Boulder, University of Illinois at Urbana-Champaign, University of Toronto, University of Utah, University of Virginia, Yale University, and Yunnan University.


\facilities{Gaia \citep{gaiamission}, GALAH \citep{galah_mission}, APOGEE \citep{apogee}, TESS \citep{Ricker15}}
\software{astropy \citep{2013A&A...558A..33A,astropyii}, scipy \citep{scipy}, pyastronomy \citep{gls}, lightkurve \citep{lightkurve}, glue \citep{glue}, astroquery \citep{astroquery}, numpy \citep{numpy}, pandas \citep{reback2020pandas,mckinney-proc-scipy-2010}, topcat \citep{topcat}, galpy \citep{galpy}, matplotlib \citep{Hunter:2007}}
\bibliographystyle{aasjournal}

\appendix 
\renewcommand\thetable{\thesection.\arabic{table}}
\renewcommand\thetable{\thesection.\arabic{figure}}
\renewcommand{\thesubsection}{\Alph{subsection}}
\counterwithin{figure}{section}
\counterwithin{table}{section}
\setcounter{figure}{0}

\section{\textbf{Injection \& Recovery Test}}\label{sec:injection}
\begin{figure}[h!]
    \centering
    \includegraphics[width=\textwidth]{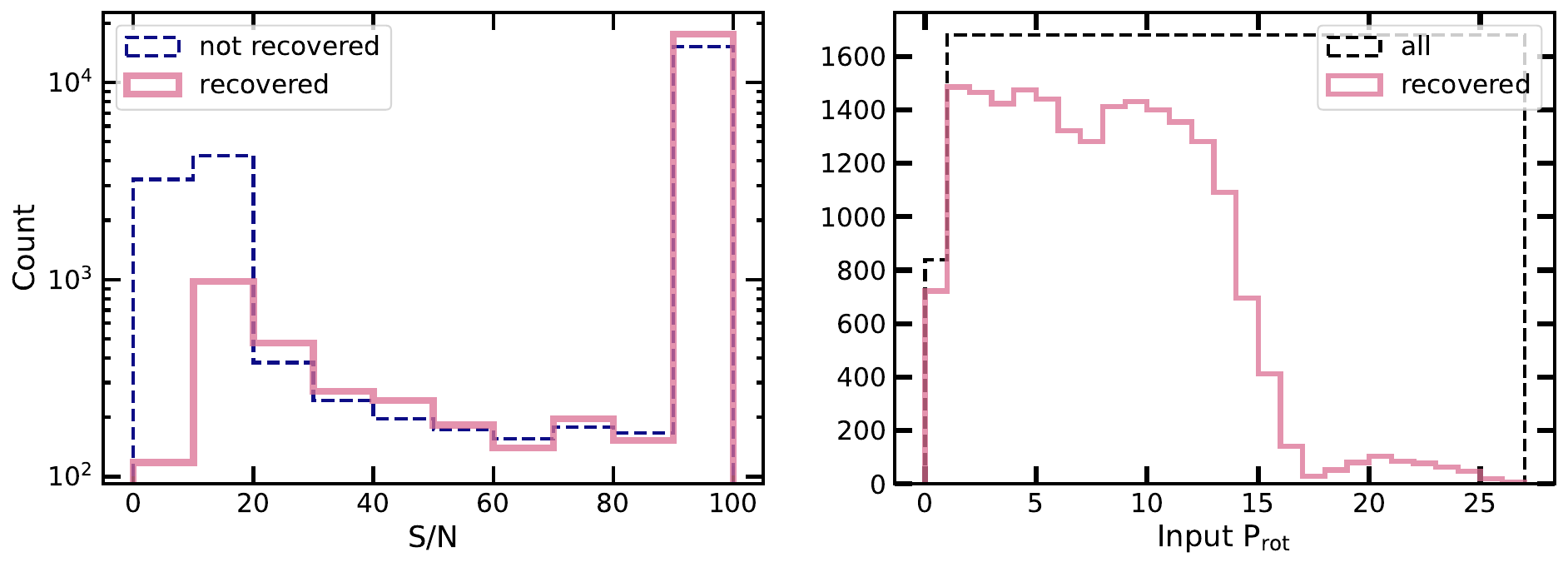}
    \includegraphics[width=\textwidth]{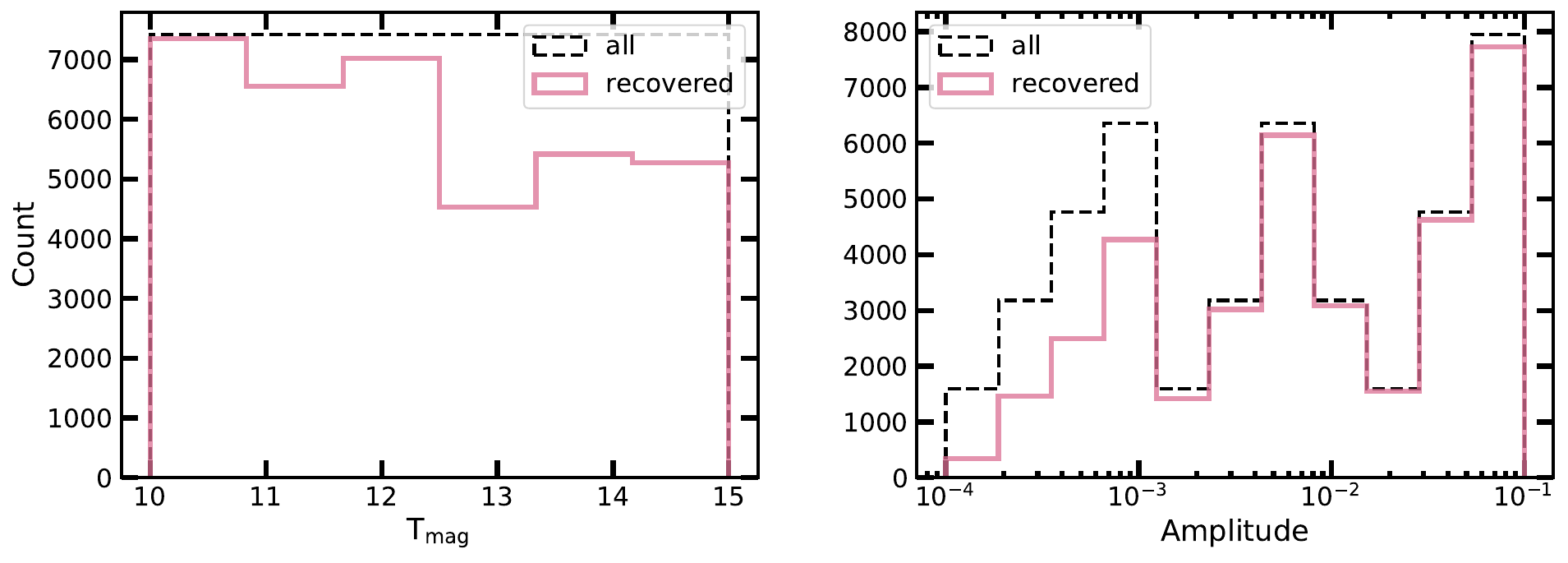}
    \caption{Distribution of stars with an injected \prot\ as a function of S/N (top left), input \prot\ (top right), magnitude (bottom left), and amplitude (bottom right) that were recovered (red solid line) versus not recovered (blue dashed line) for S/N and entire sample (black dashed line) for input \prot\, magnitude, and amplitude.}
    \label{fig:injection}
\end{figure}

\begin{figure}[h!]
    \centering
    \includegraphics[width=\textwidth]{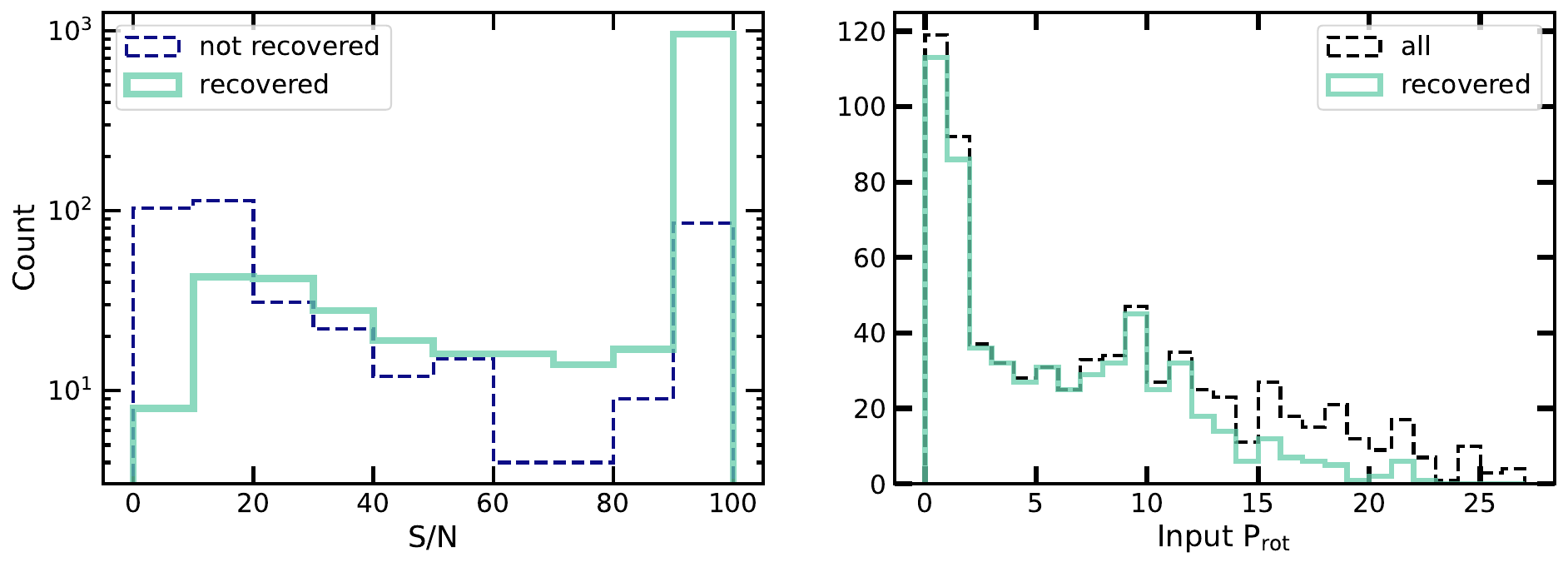}
    \includegraphics[width=\textwidth]{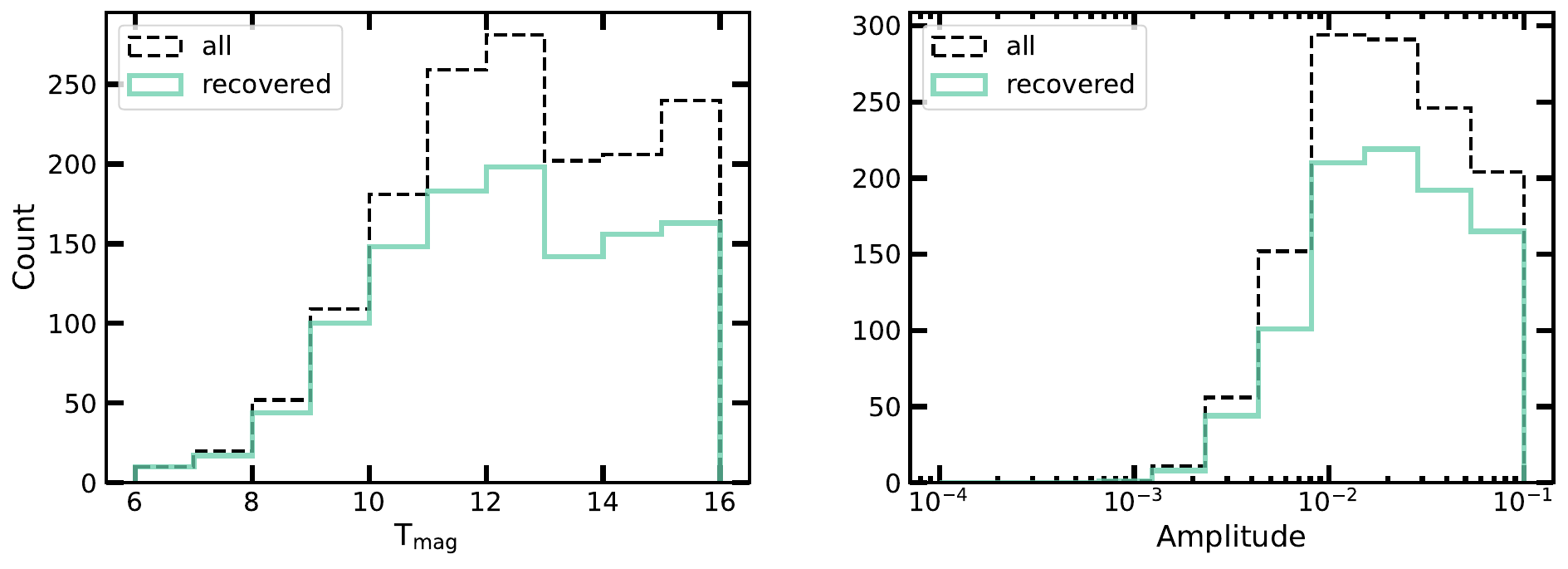}

    \caption{Distribution stars from benchmark populations as a function of S/N (top left), input \prot\ (top right), magnitude (bottom left), and amplitude (bottom right) that were recovered (green solid line) versus not recovered (blue dashed line) for S/N and entire sample (black dashed line) for input \prot\, magnitude, and amplitude.}
    \label{fig:exsector}
\end{figure}

We perform an injection and recovery test to determine I) the maximum \prot\ we can detect II) the S/N that best discriminates between astrophysical periodicity and noise. We use 25 stars in the MEarth survey with measured periods ${>}30$ days and TESS magnitudes ranging from $10-15$. We scramble the light curves and inject rotation signals comprised of a sinusoid plus the first harmonic. We inject periods from $0-27$ days in 0.5 day increments and amplitudes from 0.0001--0.001 in 0.001 increments, 0.001--0.01 in 0.001 increments, and 0.01--0.1 in 0.01 increments. The amplitude range is based on semi amplitudes we see in TESS observations of benchmark cluster rotators.

We show the results of the injection and recovery test in four histograms that show recovery as a function of S/N (top left), input \prot\ (top right), TESS magnitude (bottom left), and amplitude (bottom right) in Figure \ref{fig:injection}. Considering the top-left plot, we find that S/N of signals that are not recovered drops off steeply beyond S/N $= 20$, in contrast to those that are recovered. As seen in the top right panel, we recover the majority of injected signals with periods $<14$ days. At 13 days we recover 70\% of the signals, 41\% at 14 days, and 25\% at 15 days. Recovery significantly decreases from $> 90\%$ at the brighter end to 60-70\% between magnitudes of 12--13, as seen in the bottom right panel; this is approaching the limit of the TESS SPOC light curves (13.5). For amplitudes below 0.1\%, we can recover $\leq 70\%$ of the signals.


The signals we are injecting are idealized and non-evolving, so these results are likely best-case scenarios and will stay the same or worsen when looking at real data. Thus, we compare these results to our results from benchmark populations. Because recovery rates do not vary significantly between the samples, we combine all of the benchmark populations (Pleiades, Pisces-Eridanus, Hyades, Praesepe, and the MEarth sample). These results are shown in Figure \ref{fig:exsector}. We see similar trends as a function of S/N, input \prot, magnitude, and amplitude.

In the data, recovery as a function of magnitude also drops to 70\% beyond a magnitude of 13. Recovery as a function of amplitude remains at an average of $78 \pm 9\%$; amplitudes below 0.1\% are not present in this population. As in the injection-recovery test, the S/N of signals that are not recovered drops off beyond S/N = 20, but not as steeply. Imposing a S/N limit at 20 results results in a false negative rate of 5\%, but the trade-off is a high false positive rate of 40\%. This increases the burden of by-eye inspection by 2-3x. Therefore, we choose a S/N limit of 40, which has a more even trade off with a false positive rate of 25\% and a false negative rate of 10\%. We can also see at S/N $\approx 20$ in the top panel of Figure \ref{fig:snrexs} that the periodicity is just barely, if at all, visible by eye compared to the bottom two panels with S/N $> 40$.

As with the simulated signals, the maximum period at which we recover $> 50\%$ of the periods in the sample measurement is $ < 15$ days. At 13 days we recover 61\% of the signals (compared to the injected sample’s 70\%), 54\% at 14 days, and 44\% at 15 days. 
Because of these large decreases in recovery rates at longer periods in both the injection-recovery and benchmark population results, we set 13 days as the maximum period at which we can reliably recover periods. This is also a practical value to choose given as it is half the sector length of 27 days and corresponds to the 13.7 day orbital period of TESS.

\section{\textbf{Test Set Recovery \& Comparison Metrics}}\label{sec:testset}

\setcounter{table}{0}

\begin{deluxetable*}{ccccccccc}[h!]
\centering 
\tabletypesize{\footnotesize} 
\tablecolumns{4}
\tablewidth{0pt}
\tablecaption{Final Test Set: \prot\ pipeline versus literature \prot\ catalogs results}
\label{tbl:clusters_prot_ap_test}
\tablehead{
\colhead{Group} & 
\colhead{Pipeline} &
\colhead{\# of stars} &
\colhead{\% Recovery} & 
\colhead{\% Detection Recovery} &
\colhead {\% True Pos.} &
\colhead{\% True Neg.} & 
\colhead{\% False Pos.} & 
\colhead{\% False Neg.}
}
\startdata
 & I &45 & 100 & 98 & 98 & \nodata & \nodata & 2 \\
Pisces--Eridanus & II &43 & 100 & 100 & 100 & \nodata & \nodata & 0 \\
& III &45 & 100 & 98 & 98 & \nodata & \nodata & 2 \\
\\
& I & 208 & 80 & 71 &86 & 90 & 10 & 14 \\
Praesepe & II & 101 & 80 & 68 & 83 & 96 & 4 & 17 \\
& III & 208 &80 & 68 & 81 & 98 & 2 & 19 \\
\\
& I & 80 & 95 & 93 & 93 &70 & 30 & 7 \\
MEarth & II & 52 & 95 & 96 & 96 & 86 & 14 & 4 \\
 & III & 80 & 95 & 93 & 93 & 93 & 7 & 7 \\
\enddata
\tablecomments{Percentage recovered indicates whether 1, 2, or 1/2 times the literature \prot\ was recovered with each pipeline within 2x the peak-width and within the reliable \prot\ detection limit ($\leq 13$ days). Percentage of detections recovered indicates whether 1, 2, or 1/2 times the literature \prot\ was recovered with each pipeline for a detection within 2x the peak-width and within the reliable \prot\ detection limit ($\leq 13$ days). True positive indicates a detection within the reliable \prot\ detection limit. True negative indicates a non-detection outside of the the reliable \prot\ detection limit. False positive indicates a detection outside of the the reliable \prot\ detection limit. False negative indicates a non-detection within the reliable \prot\ detection limit. The Pleiades and Pisces--Eridanus do not have true negative and false positive columns since $\leq 2$ stars have \prot\ $> 13$ days. Pipeline II only considers stars with multiple sectors.}
\end{deluxetable*}



\end{document}